%% file: draft_Lc_xi0kpi0.tex
\documentclass[prd,twocolumn,showpacs,amsmath,amssymb]{revtex4-2}
\usepackage[mathlines]{lineno}
\usepackage{overpic,graphicx}
\usepackage{dcolumn}
\usepackage{bm}
\usepackage{rotating}
\usepackage{subfigure}
\usepackage{color}
\usepackage{multirow}
\usepackage{epstopdf}
\usepackage{mathrsfs}
\usepackage{comment}
\usepackage{booktabs}
\usepackage{hyperref}
\usepackage{define}

\let\oldequation\equation
\let\oldendequation\endequation

\renewenvironment{equation}
  {\linenomathNonumbers\oldequation}
  {\oldendequation\endlinenomath}
  
\begin{document}

\title{Measurement of the absolute branching fraction of the three-body decay $\LamC \to \Xi^{0}K^{+}\pi^{0}$ and search for $\LamCNKPi{}$, $\Sigma^{0}K^{+}\pi^{0}$ and $\Lambda K^{+}\pi^{0}$
}

\input{authorlist_2023-3-6.tex}

\begin{abstract}

The Cabbibo-favored decay $\LamCXiKPi{}$ is studied for the first time
using 6.1 $\ifb{}$ of $\ee{}$ collision data at center-of-mass
energies between 4.600 and 4.840 GeV, collected with the BESIII
detector at the BEPCII collider. With a double-tag method, the
branching fraction of the three-body decay $\LamCXiKPi{}$ is measured
to be $(7.79 \pm 1.46 _{\rm} \pm0.95 _{\rm}) \times 10^{ - 3}$, where
the first and second uncertainties are statistical and systematic,
respectively. The branching fraction of the two-body
decay~$\LamCXistarK{}$~is $(5.99\pm1.04\pm0.32)\times
10^{-3}$, which is consistent with the previous result
of $(5.02\pm0.99\pm0.31)\times 10^{-3}$. In addition, the upper limit on
the branching fraction of the doubly Cabbibo-suppressed decay
$\LamCNKPi{}$ is $7.1 \times 10^{-4}$ at the 90$\%$ confidence
level. The upper limits on the branching fractions of $\LamC \to
\Sigma^{0}K^{+}\pi^{0}$ and $\Lambda K^{+}\pi^{0}$ are also determined
to be $1.8\times 10^{-3}$ and $ 2.0 \times 10^{-3}$, respectively.

\end{abstract}


\maketitle

\oddsidemargin  -0.2cm
\evensidemargin -0.2cm

\section{Introduction}

\begin{figure*}[!htp]
\centering
\subfigure[]{
\includegraphics[height=0.103\textheight]{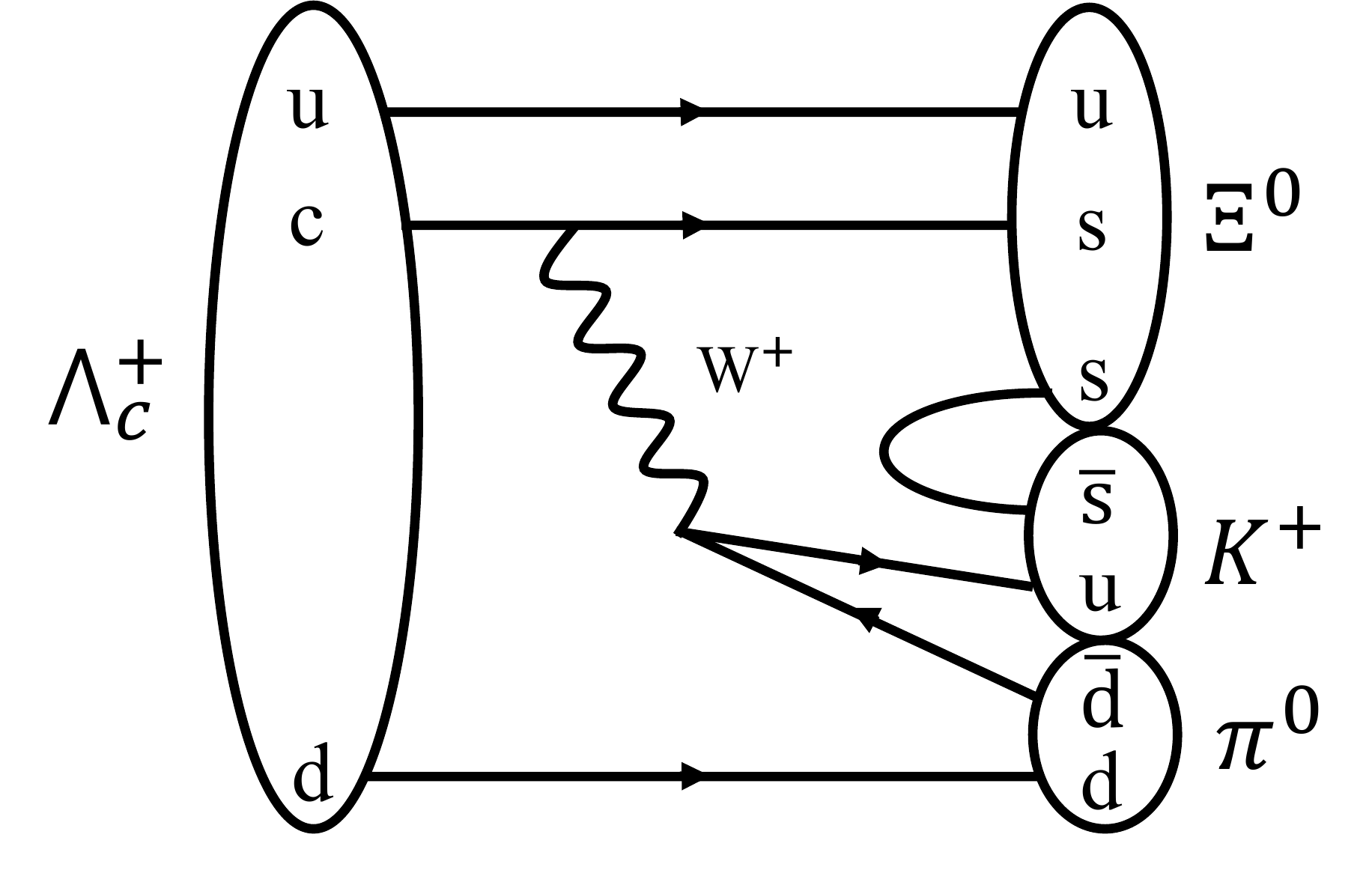}\label{fig:fey1}}
 \subfigure[]{
 \includegraphics[height=0.103\textheight]{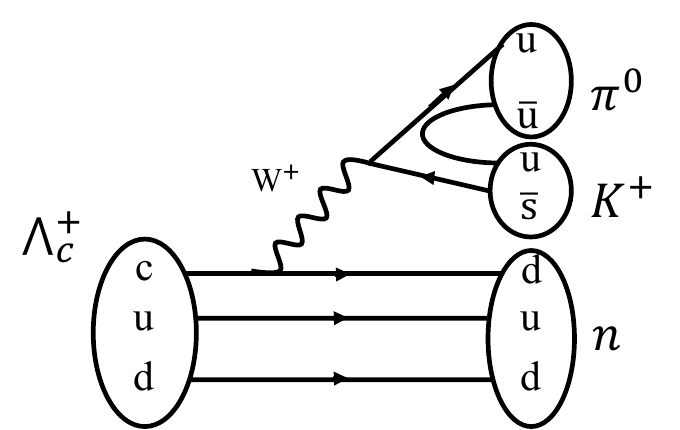}\label{fig:fey3}}
 \subfigure[]{
\includegraphics[height=0.103\textheight]{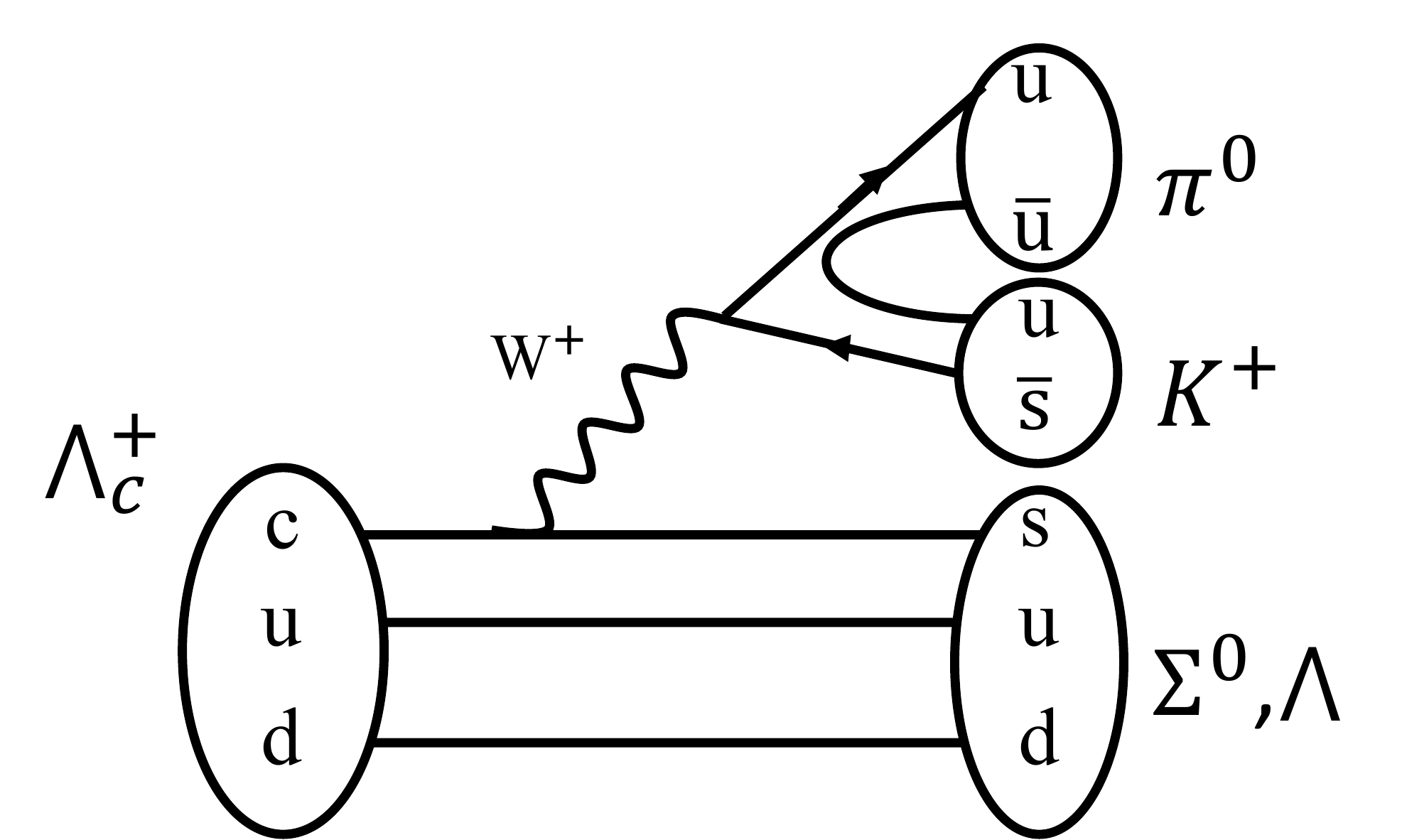}\label{fig:fey2}}
\subfigure[]{
\includegraphics[height=0.103\textheight]{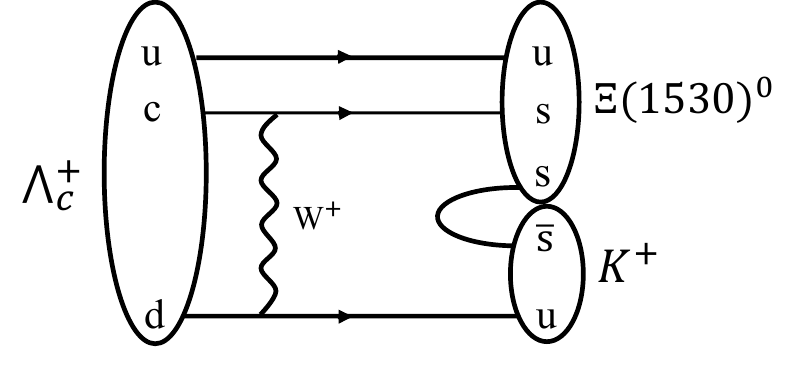}\label{fig:fey4}}
\caption{Typical Feynman diagrams of the decay (a) \LamCXiKPi{}, (b) $nK^{+}\pi^{0}$, (c) $\Sigma^{0}(\Lambda)K^{+}\pi^{0}$ and (d) $\Xi(1530)^{0}K^{+}$. }\label{fig:fey}
\end{figure*}

Hadronic decays of charmed baryons provide an ideal laboratory to
understand the interplay of weak and strong interactions in the charm
system~\cite{01,02,03,04,05,06,07,08,09}, and the measured branching
fractions of charmed baryons play an important role in constraining
models of charmed hadronic decays~\cite{model1,model2}.  However, no
reliable phenomenological model of charmed baryon decays currently
exists. The \LamC{} baryon is the lightest charmed baryon, thus it is
often a decay product in many charmed baryon decays. Hence, an
investigation of the \LamC{} decay is essential for understanding
excited charmed baryons~\cite{z34}.  In charmed baryon decays, the
non-factorizable contributions from W-exchange diagrams play an
essential role. In contrast, these effects are negligible in heavy
meson decays~\cite{z27}.  Therefore, measurements of
the absolute branching fractions of \LamC{} decays are important to
understanding the internal dynamics of charmed baryon
decays~\cite{z12}.

In recent years, great progress has been made in the experimental
study of the \LamC{} baryon at the Belle~\cite{z35}, LHCb~\cite{LHCb}
and BESIII experiments~\cite{z13, z14, z15, z36}. First, a
breakthrough was achieved with the measurement of the absolute
branching fraction of $\LamCPKPi$~\cite{z28,z33}.  Next, BESIII
directly measured the absolute branching fractions of twelve decay
modes for the first time~\cite{z28}.  However, precise measurements
for several Cabbibo-favored (CF) decays of $\Lambda^+_c$ are currently
unavailable. Furthermore, a large number of the singly
Cabbibo-suppressed (SCS) decays and the doubly Cabbibo-suppressed
(DCS) decays of $\Lambda^+_c$ have not yet been studied. Based on
the data sets at BESIII, measurements of the branching fractions of
$\Lambda^+_c$ decays of currently known decay modes could potentially
reach a sensitivity of $10^{-4}$. In particular, the branching
fraction of $\Lambda_{c}^+ \to n \pi^+$ was recently measured to be
$(6.6\pm1.2\pm0.4)\times 10^{-4}$~\cite{z15}. In addition, the data allows
for a search of many rare $\Lambda^+_c$ decays.

The three-body CF decay \LamCXiKPi{} is expected to have a large decay
rate. Figure~\ref{fig:fey1} presents a typical Feynman diagram of this
decay. Some phenomenological models have predicted different branching
fractions for the \LamCXiKPi{} decay, which are $(4.5\pm0.8) \times
{10^{ - 2}}$~\cite{z16} and $(3.2\pm0.6) \times {10^{ -
2}}$~\cite{z30} by assuming SU(3) flavor symmetry and isospin
asymmetry, respectively. Experimentally, there are only a few studies
of \LamC{} decays with a $\Xi^{0}$ baryon in the final state, and the
three-body decay \LamCXiKPi{} has not yet been studied. Measurement of
the branching fraction will help us further understand the underlying
dynamics of $\Lambda^+_c$ decays and distinguish among the different
theoretical models~\cite{z16}.

The three-body DCS decay \LamCNKPi{} is also of great interest, and
phenomenological models of this decay mode have been
proposed~\cite{z16, z30}. SU(3) flavor symmetry predicts the branching
fraction of \LamCNKPi{} to be $(5.0\pm0.5) \times {10^{ -
5}}$~\cite{z16}.  Figure~\ref{fig:fey3} shows a typical Feynman
diagram of \LamCNKPi{}.  Experimental measurements of DCS decays of
$\Lambda^+_c$ are limited due to the suppressed branching
fractions. Currently, the only study of a DCS \LamC{} decay is
${\ensuremath{\LamC\to pK^+\pi^-}}$. The relative branching fraction
${\mathcal B}({\ensuremath{\LamC\to pK^+\pi^-}})$/ ${\mathcal
B}({\ensuremath{\LamC\to pK^-\pi^+}})$ is measured to be $(2.4 \pm
0.3 \pm 0.2)\times 10^{-3}$~\cite{z32}.

In this paper, the three-body decays \LamCXiKPi{}, \LamCNKPi,
\LamCSigKPi, \LamCLamKPi, and the two-body decay \LamCXistarK{} are
studied experimentally with a double-tag (DT) method~\cite{z39}.  Figure~\ref{fig:fey2} and~\ref{fig:fey4} show typical Feynman
diagrams of ${\ensuremath{\LamC\to \Sigma^{0}  (\Lambda)K^+\pi^0}}$ and $\LamCXistarK{}$, respectively. 
This analysis is performed using a data sample with an integrated
luminosity of 6.1 \ifb{} collected at center-of-mass (CM) energies
between 4.600 and 4.840 GeV, listed in Table~\ref{tab:data_sets}, by
the BESIII detector at the BEPCII collider. Throughout the text, the
charge conjugate states are always implied.

\begin{table}[!htbp]
  \begin{center}
    \caption{The CM energy and the integrated luminosity ($\mathcal{L}_{\rm int}$)
         for each energy point.
             The first and the second uncertainties are statistical and systematic, respectively.}
    \begin{tabular}{c| c c}
      \hline
      \hline
    Dataset    & CM energy (MeV) &  $\mathcal L_{\rm int}$ (\ipb)  \\
      \hline
        4.600    &  4599.53 $\pm$ 0.07 $\pm$ 0.74   &  586.90  $\pm$ 0.10 $\pm$ 3.90 \\
        4.612   & 4611.84 $\pm$ 0.12 $\pm$ 0.28   &  103.45  $\pm$ 0.05 $\pm$ 0.64 \\
        4.620   & 4628.00 $\pm$ 0.06 $\pm$ 0.31   &  519.93  $\pm$ 0.11 $\pm$ 3.22 \\
        4.640    & 4640.67 $\pm$ 0.06 $\pm$ 0.36   &  548.15  $\pm$ 0.12 $\pm$ 3.40 \\
        4.660    & 4661.22 $\pm$ 0.06 $\pm$ 0.29   &  527.55  $\pm$ 0.12 $\pm$ 3.27 \\
        4.680    & 4681.84 $\pm$ 0.08 $\pm$ 0.29   &  1664.34 $\pm$ 0.21 $\pm$ 10.32 \\
        4.700    & 4698.57 $\pm$ 0.10 $\pm$ 0.32   &  534.40  $\pm$ 0.12 $\pm$ 3.31 \\
        4.740   & 4739.70 $\pm$ 0.20 $\pm$ 0.30   &  164.27  $\pm$ 0.07 $\pm$ 0.87 \\
        4.750    & 4750.05 $\pm$ 0.12 $\pm$ 0.29   &  367.21  $\pm$ 0.10 $\pm$ 1.95 \\
        4.780    & 4780.54 $\pm$ 0.12 $\pm$ 0.33   &  512.78 $\pm$ 0.12 $\pm$ 2.72 \\
        4.840    & 4843.07 $\pm$ 0.20 $\pm$ 0.31   &  527.29  $\pm$ 0.12 $\pm$ 2.79 \\
      \hline\hline
    \end{tabular}
      \label{tab:data_sets}
  \end{center}
\end{table}

\section{BESIII detector and Monte Carlo} The BESIII
detector~\cite{BESCol} records symmetric $e^+e^-$ collisions provided
by the BEPCII~\cite{BEPCII} storage ring, which operates in the CM
energy range from 2.0 to 4.95~GeV, with a peak luminosity of $1.1
\times 10^{33}\;\text{cm}^{-2}\text{s}^{-1}$ achieved at $\sqrt{s} =
3.77\;\text{GeV}$~\cite{detector1}.  The cylindrical core of the
BESIII detector covers 93\% of the full$~$ solid angle and consists of
a helium-based multilayer drift chamber~(MDC), a plastic scintillator
time-of-flight system~(TOF), and a CsI(Tl) electromagnetic
calorimeter~(EMC), which are all enclosed in a superconducting
solenoidal magnet providing a {\spaceskip=0.2em\relax 1.0 T} magnetic
field. The solenoid is supported by an octagonal flux-return yoke with
resistive plate counter muon identification modules interleaved
with steel~\cite{detectorY}.  The charged-particle momentum resolution
at $1~{\rm GeV}/c$ is $0.5\%$, and resolution of the ionization energy
loss in the MDC ($\mathrm{d}E/\mathrm{d}x$) is $6\%$ for electrons
from Bhabha scattering. The EMC measures photon energies with a
resolution of $2.5\%$ ($5\%$) at $1$~GeV in the barrel (end-cap)
region.  The time resolution in the TOF barrel region is 68~ps.  The
end-cap TOF system was upgraded in 2015 using multi-gap resistive
plate chamber technology, providing a time resolution of 60
ps~\cite{detector2}.

Simulated samples produced with {\sc geant4}-based~\cite{geant4} Monte
Carlo software, which includes the geometric description~\cite{geod1,
geod2} of the BESIII detector and the detector response, are used to
determine the detection efficiency and to estimate backgrounds. Exclusive simulation samples of $e^+e^- \to \Lambda^{+}_c
\bar{\Lambda}^{-}_c$ are produced with $\bar{\Lambda}^{-}_c$ decaying
into ten specific tag modes and $\LamC$ decaying into
$\Xi^{0}K^{+}\pi^{0}$ and $nK^{+}\pi^{0}$.  
The resonances are modeled with Breit-Wigner functions, of which masses and widths are taken from the Particle Data Group (PDG)~\cite{g3}. 
The simulation includes
the beam-energy spread and initial-state radiation (ISR) in the
$e^+e^-$ annihilations with the specific tag modes~\cite{peta} modeled
with the generator {\sc kkmc}~\cite{ref:kkmc1, ref:kkmc2}. The signal
events are modeled with a phase-space generator. The $e^+e^- \to
\Lambda^{+}_c \bar{\Lambda}^{-}_c$ line-shape implements the
description from Ref.~\cite{z13}.  The inclusive simulation sample,
which consists of $\Lambda_c^+\bar{\Lambda}_c^-$ events, $D_{(s)}$
production,$~$ISR return to lower-mass $\psi$ states, and continuum
processes ($e^{+}e^{-}\rightarrow u\bar{u}, d\bar{d}$ and $s\bar{s}$)
is used to estimate the potential background. All the known decay
modes of charmed hadrons and charmonia are modeled with {\sc
evtgen}~\cite{g1, g2} using branching fractions either taken from the
PDG~\cite{g3}, when available, or otherwise estimated with {\sc lundcharm}~\cite{g4, g4a}.  Final state radiation
from charged final state particles is incorporated using {\sc
photos}~\cite{g5}.

\section{Methodology}

The $\LamCB$ baryons are fully reconstructed by their hadronic decays
to $ \bar {p}K^+\pi^-$, $\bar {p}K^{0}_{S}$, $\bar {p} K^+ \pi^-
\pi^0$, $\bar {p}K^{0}_{S}\pi^0$, $\bar {p}K^{0}_{S}\pi^+\pi^-$,
$\bar\Lambda\pi^-$, $\bar\Lambda\pi^-\pi^-\pi^+$,
$\bar\Lambda\pi^-\pi^0$, $\bar{\Sigma}^0\pi^-$ and
$\bar{\Sigma}^-\pi^+\pi^-$. These reconstructed decays are referred to as
single-tag (ST) $\bar\Lambda^{-}_{c}$ baryons, where the intermediate
particles $K^{0}_{S}$, $\bar{\Lambda}$, $\bar{\Sigma}^0$,
$\bar{\Sigma}^-$, and $\pi^0$ are reconstructed via $K^{0}_{S}\to
\pi^+\pi^-$, $\bar{\Lambda}\to\bar{p}\pi^+$,
$\bar{\Sigma}^0\to\gamma\bar{\Lambda}$,
$\bar{\Sigma}^-\to\bar{p}\pi^0$ and $\pi^0\to\gamma\gamma$,
respectively.  From the remaining tracks and showers, the three-body
decays $\LamCXiKPi{}$, $nK^{+}\pi^{0}$, $\Sigma^{0}K^{+}\pi^{0}$ and $\Lambda K^{+}\pi^{0}$ are selected to form
$\Lambda_c^+$ candidates, which will be referred to as the recoiling
system. Together with the $\bar\Lambda^{-}_{c}$ candidates, these form
a sample of DT events.

The branching fraction
is expressed as:

\begin{equation} \label{eq:br1} {\mathcal B} = \frac{N_{\mathrm{sig}}}
{{\mathcal B}_{\mathrm{inter}} \cdot \Sigma_{i}\, N^{\rm ST}_{i} \cdot
(\epsilon^{\rm DT}_{i} / \epsilon^{\rm ST}_{i}) }, \end{equation}

\noindent where $N_{\mathrm{sig}}$ is the signal yield of the DT
candidates and ${\mathcal B}_{\mathrm{inter}}$ is the product of
branching fractions of intermediate decays on
the signal side from the PDG~\cite{g3}. $N^{\rm ST}_{i}$ denotes the yields of the ST candidates observed in data, and $\epsilon^{\rm ST}_{i}$ and
$\epsilon^{\rm DT}_{i}$ are the ST and DT efficiencies, respectively.

\section{Event selection}

The selection criteria in this analysis follow that of
Ref.~\cite{z15}. Charged tracks are required to have a polar angle
($\theta$) within $|\!\cos\theta| < 0.93$, where $\theta$ is defined
with respect to the $z$-axis, which is the symmetry axis of the
MDC. Tracks, except for those from $K^0_S$ and $\bar{\Lambda}$ decays,
are required to originate from an interaction region defined by
$|V_{xy}| < 1$~cm and $|V_z| < 10$~cm (referred as a tight track
hereafter), where $|V_{xy}|$ and $|V_z|$ refer to the distances of
closest approach of the reconstructed track to the interaction point
(IP) in the $xy$ plane and the $z$ direction, respectively.

Particle identification~(PID) for charged tracks combines measurements of the energy deposited in the MDC~(d$E$/d$x$) and the flight time in the TOF to form likelihoods $\mathcal{L}(h)~(h=p,K,\pi)$ for each hadron $h$ hypothesis.
Tracks are identified as protons when the proton hypothesis has the greatest likelihood ($\mathcal{L}(p)>\mathcal{L}(K)$ and $\mathcal{L}(p)>\mathcal{L}(\pi)$), while charged kaons and pions are identified by comparing the likelihoods for the kaon and pion hypotheses, $\mathcal{L}(K)>\mathcal{L}(\pi)$ and $\mathcal{L}(\pi)>\mathcal{L}(K)$, respectively.

The selection of tracks from $K^{0}_{S}$ and $\bar{\Lambda}$ is
different from those of other tracks. Candidates of $K^{0}_{S}$ and
$\bar{\Lambda}$ hadrons are reconstructed from their decays to
$\pi^+\pi^-$ and $\bar{p}\pi^+$, respectively, where the charged
tracks must satisfy $|V_z| < 20$ cm (referred as a loose track
hereafter).  The PID selection criteria
are imposed on the antiproton candidate, while the charged pion is not
subject to any PID requirement.  A secondary vertex fit is performed
for each $K^{0}_{S}$ or $\bar{\Lambda}$ candidate, and the momentum
obtained from the fit is used in the subsequent analysis. The
$K^{0}_{S}$ or $\bar{\Lambda}$ candidate is retained if the $\chi^2$
of the secondary vertex fit is less than 100.  Furthermore, the decay
vertex is required to be separated from the IP by a distance of at
least twice the vertex resolution. For $K^{0}_{S}$ and
$\bar{\Lambda}$, the invariant masses of $\pi^+\pi^-$ and $\bar
p\pi^+$ pairs are required to be within (0.487, 0.511)~GeV/$c^2$ and
(1.111, 1.121)~GeV/$c^2$, respectively. The $\pi^+\pi^-$ and
$\bar{p}\pi^-$ invariant mass resolutions determined through
simulations, are 2.9 MeV$/c^{2}$ and 1.2 MeV$/c^{2}$,
respectively. Similarly, the $\bar{\Sigma}^0$ and $\bar{\Sigma}^-$ candidates are
reconstructed from the $\gamma\bar{\Lambda}$ and $\bar{p}\pi^0$ final
states with invariant masses within the ranges of (1.179, 1.203)~GeV/$c^2$ and
(1.176, 1.200)~GeV/$c^2$, respectively. The mass resolutions for $\bar \Sigma^0$ and $\bar
\Sigma^-$ are found, using simulation, to be 3.6
MeV$/c^{2}$ and 4.3 MeV$/c^{2}$, respectively.

Photon candidates are identified using showers in the EMC. The
deposited energy of each shower must be more than 25~MeV in the barrel
region ($|\!\cos\theta| \le 0.80$) or more than 50~MeV in the end-cap
region ($0.86 \le |\!\cos\theta| \le 0.92$). To suppress electronic
noise and showers unrelated to the event, the difference between the
EMC time and the event start time is required to be within (0,
700)~ns.

A $\pi^0$ candidate is reconstructed with a photon pair, and their
invariant mass is required to be within the range (0.115,
0.150)~GeV/$c^2$. To improve the resolution, a kinematic fit is
performed, where the diphoton invariant mass is constrained to the
known $\pi^0$ mass~\cite{g3}, and the $\chi^2$ of the kinematic fit is
required to be less than 200. The momenta obtained from the kinematic
fit are used in the subsequent analysis.

To distinguish the ST $\bar\Lambda^{-}_{c}$ baryons from combinatorial
backgrounds, the distributions of the energy difference $\Delta E$ are
used, defined as 

\begin{equation} \Delta E\equiv
 E_{\bar\Lambda^{-}_{c}}-E_{\mathrm{beam}}, 
\end{equation} 

\noindent where $E_{\rm beam}$ is the beam energy, and
 $E_{\bar\Lambda^{-}_{c}}$ is the total energy of the ST candidate,
 calculated in the $e^+e^-$ rest frame. The signal events are expected
 to peak around zero in the $\Delta E$ distribution.  If an event has
 multiple $\bar\Lambda^{-}_{c}$ candidates, the one with the smallest
 $|\Delta E|$ is retained, and the $\Delta E$ requirements are listed in
 Table~\ref{tab:yield-st-460}.

The beam-constrained mass $M_{\rm BC}$ of the selected ST candidates is
defined as

\begin{equation}
M_{\rm BC}\equiv\sqrt{E_{\mathrm{beam}}^{2}/c^{4}-|\vec{p}_{\bar\Lambda^{-}_{c}}|^{2}/c^{2}},
\end{equation}

\noindent where $\vec{\mkern1mu p}_{\bar\Lambda^{-}_{c}}$ is the total
momentum of the ST candidate and $M_{\rm BC}$ peaks at the $\LamCB$
mass.  For each tag mode, the ST yield is determined by fitting the
$M_{\rm BC}$ distribution. In the fit, the $\LamCB$ signal is described by
the MC simulated shape convolved with a Gaussian function to
compensate for the resolution difference between data and MC. The
parameters of the Gaussian function are free in the fit and different
for each energy and tag mode. The combinatorial background is
described by an ARGUS function~\cite{ARGUS}, with the end-point
parameter fixed to the nominal beam energy.  The fits to the $M_{\rm
BC}$ distributions of the various tag modes for the 4.600 - 4.700 datasets
are the same as Ref.~\cite{peta}.  The fits to the $M_{\rm BC}$
distributions for the 4.740, 4.750, 4.780, and 4.840 datasets are shown
in Fig.~\ref{fig:single-tag}. Candidates in the $M_{\rm BC}$ signal
region, $(2.275, 2.310)$~GeV$/c^2$, are kept for further analysis. The
ST yields in data and the ST efficiencies for individual tags are
listed in Table~\ref{tab:yield-st-460}. The total ST yield, obtained by
summing the ST yield of all tag modes and all energies, is found to be
$N^{\rm tot}_{\rm ST}=130439 \pm 425$, where the uncertainty is
statistical.

 \begin{figure*}[!htp]
    \centering
    \setlength{\abovecaptionskip}{-0.06 cm}
             \subfigure[]{\includegraphics[width= 0.42\textwidth]{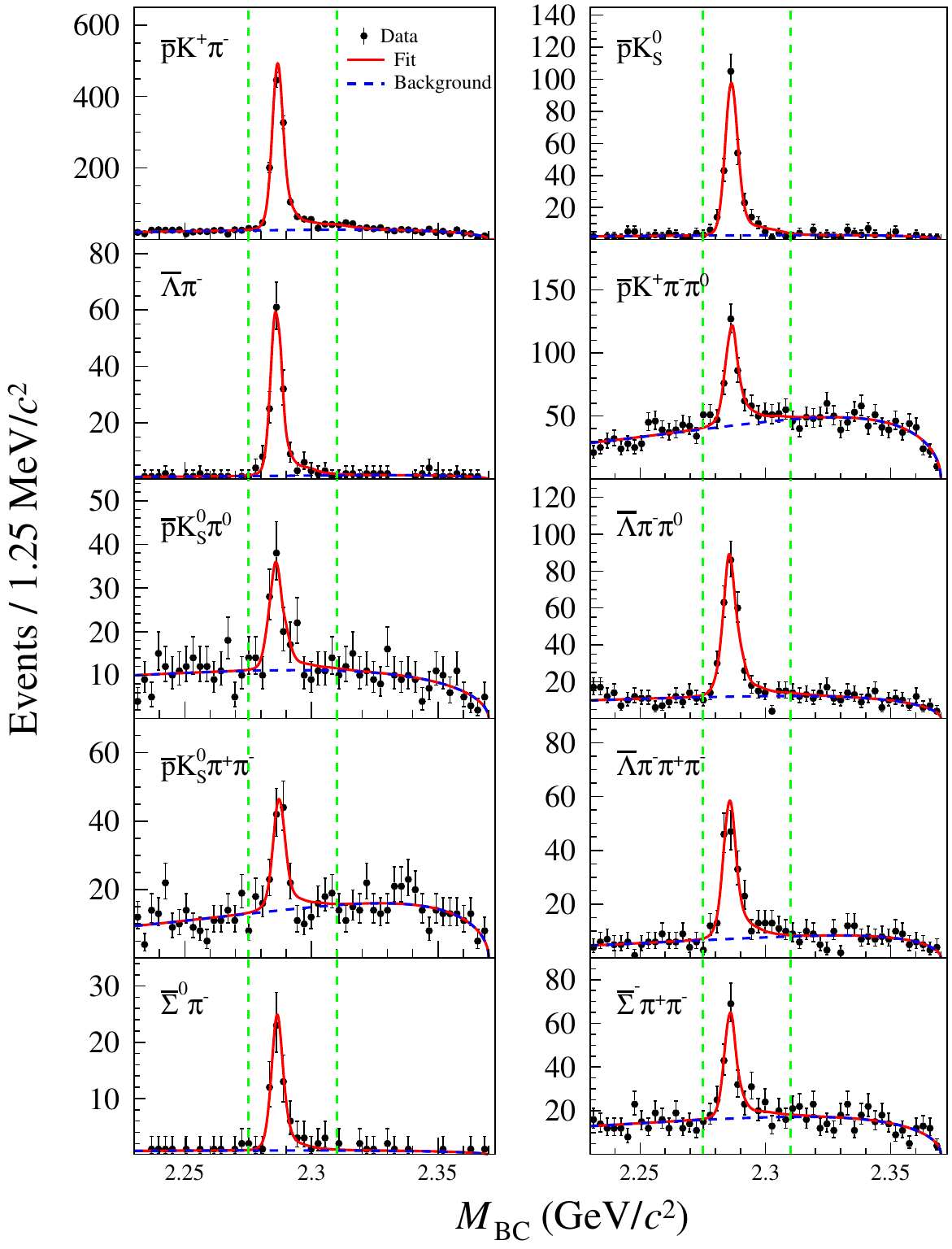}\label{fig:single-tag-474}}
               \hspace{3em}%
            \subfigure[]{\includegraphics[width= 0.42\textwidth]{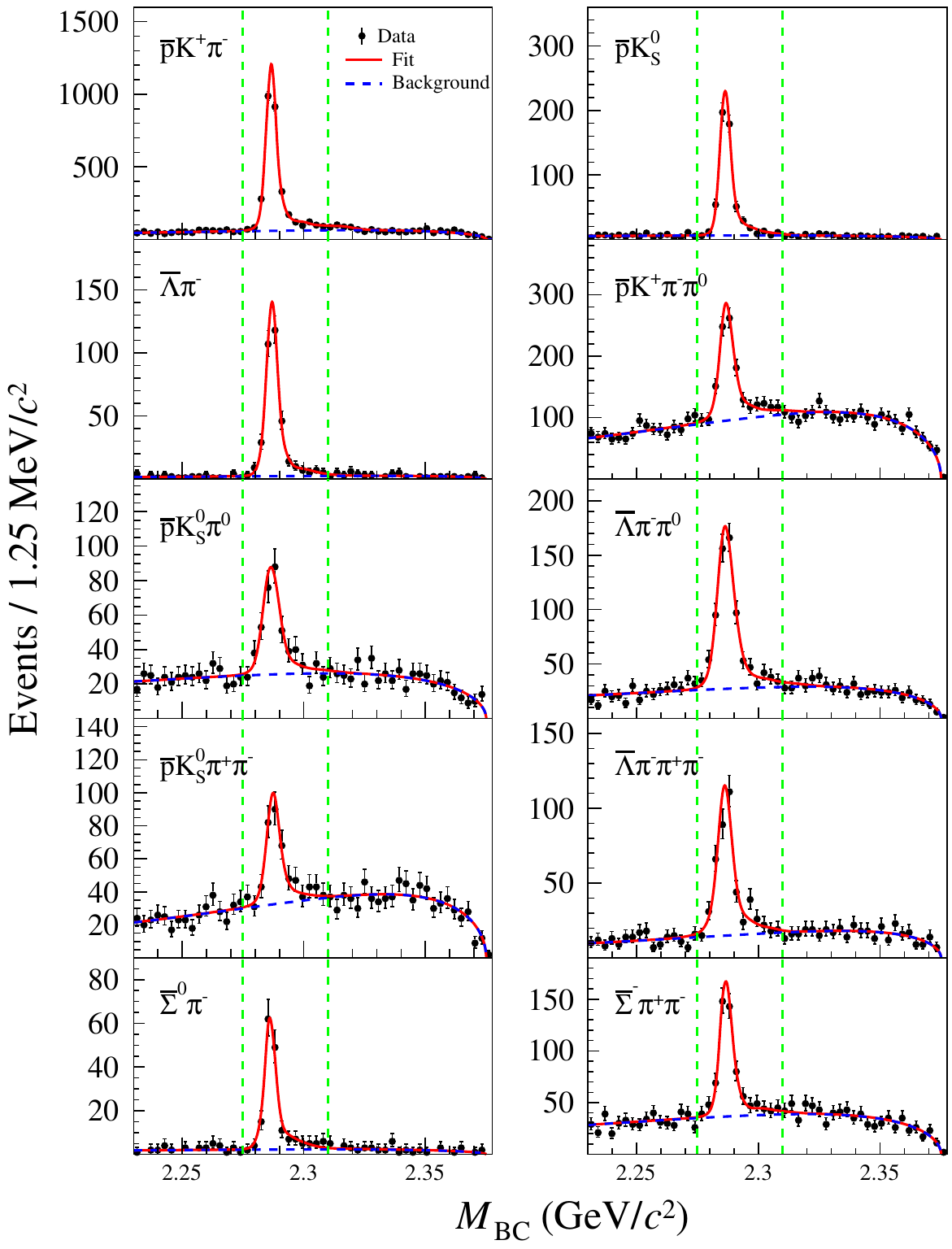}\label{fig:single-tag-475}}
            \subfigure[]{\includegraphics[width= 0.42\textwidth]{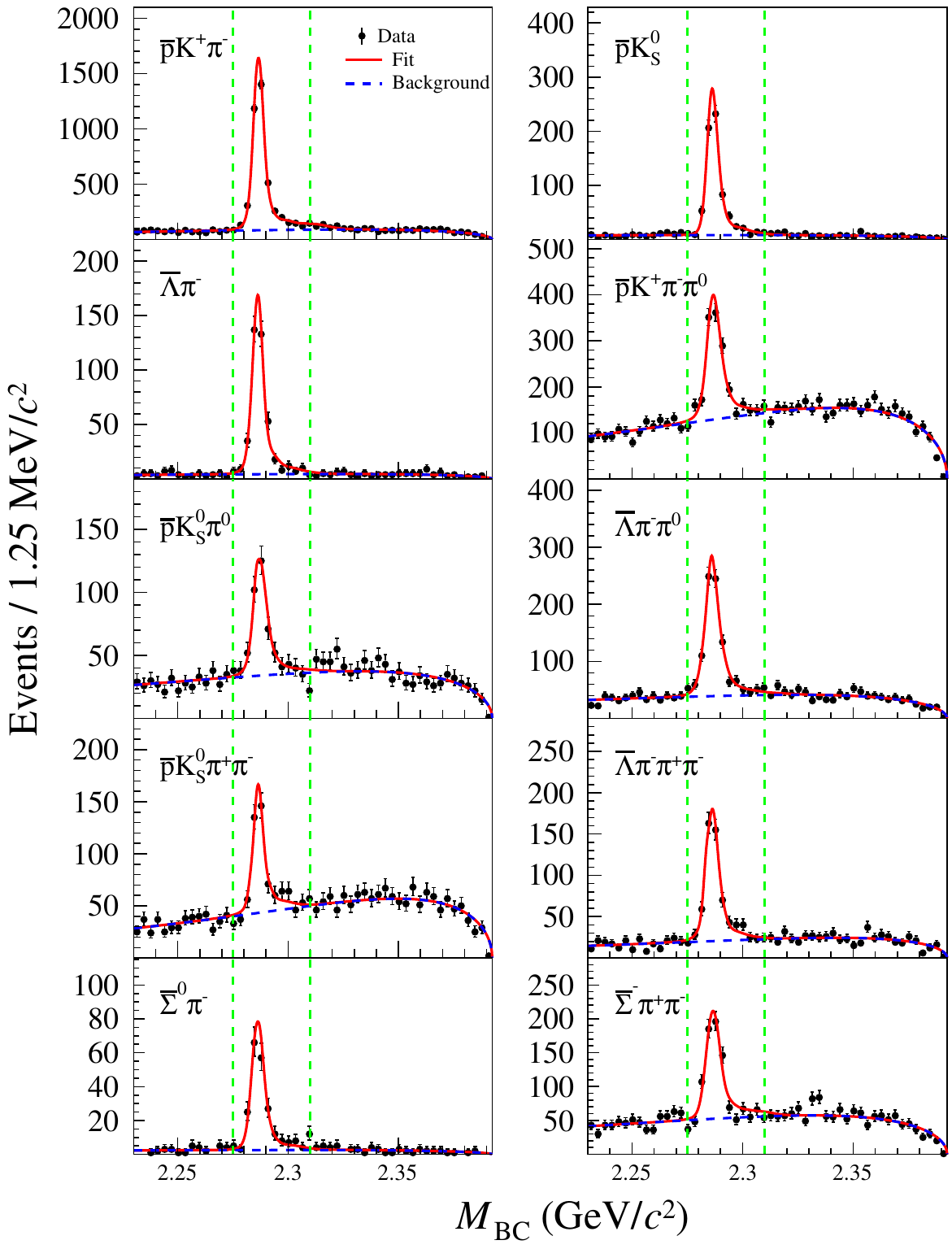}\label{fig:single-tag-478}}
              \hspace{3em}%
            \subfigure[]{\includegraphics[width= 0.42\textwidth]{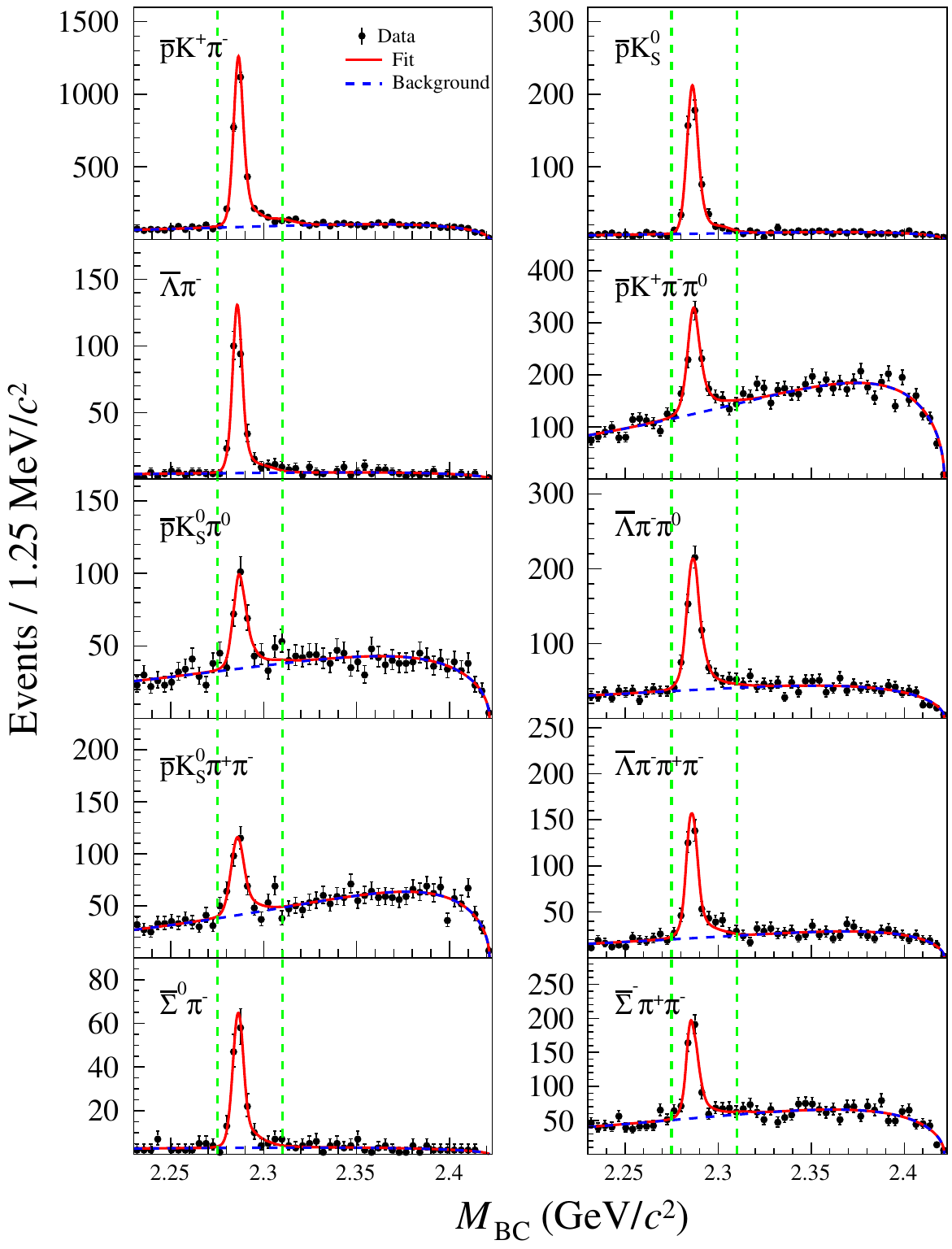}\label{fig:single-tag-484}}
        \caption{ The $\mbc$ distributions of ST channels of $\LamCB$ at (a) 4.740 GeV, (b) 4.750 GeV, (c) 4.780 GeV and (d) 4.840 GeV.
      The signal shape of the $\LamCB$ is described
      by the simulated shape convolved with a Gaussian resolution
      function, and the background is modeled
      with an ARGUS function. The points with error bars represent
data. The (red) solid curves indicate the fit results and the
(blue) dashed curves describe the background shapes. The vertical
dashed lines show the $\mbc$ selection requirements. }\label{fig:single-tag}
      \end{figure*}

\begin{table*}[!htbp]
\renewcommand\arraystretch{1.15}
  \begin{center}
  \caption{The $\Delta E$ requirements, the ST yields, and the ST detection efficiencies 
  of 
each tag mode for the data samples between 4.740 and 4.840 $\gev$. The uncertainties of the ST yields are statistical only. }
    \begin{tabular}{ l c r @{ $\pm$ } p{0.3 cm}c    r @{ $\pm$ } p{0.3 cm}c   r @{ $\pm$ } p{0.3 cm}c     r @{ $\pm$ } p{0.3 cm}c c}
      \hline
      \hline
   &   &  \multicolumn{3}{c}{4.740} &  \multicolumn{3}{c}{4.750} &  \multicolumn{3}{c}{4.780}  & \multicolumn{3}{c}{4.840} \\ 
      \hline        
    Tag mode & $\Delta E$ (MeV) & \multicolumn{2}{c}{$N_{i}^{\mathrm{ST}}$}  & $\epsilon_{i}^{\mathrm{ST}}$(\%)   & \multicolumn{2}{c}{$N_{i}^{\mathrm{ST}}$}  & $\epsilon_{i}^{\mathrm{ST}}$(\%)  & \multicolumn{2}{c}{$N_{i}^{\mathrm{ST}}$}  & $\epsilon_{i}^{\mathrm{ST}}$(\%)  & \multicolumn{2}{c}{$N_{i}^{\mathrm{ST}}$}  & $\epsilon_{i}^{\mathrm{ST}}$(\%)  \\
     \hline
 
            $\Bpkpi$                                  & $(-34,~20)$    &  $1201$   & 40  &  48.4   & 2782  &  60  & 48.1  &  3689 & 67  &  47.1&  2676 & 61  &  43.8\\
            $\Bpks$                                   & $(-20,~20)$    &  $256$     & 17  &  48.5   & 525  &  24  & 46.9  &  629 & 26  &  44.2 &  475 & 23  &  42.4\\
            $\bar{\Lambda}\pi^-$               & $(-20,~20)$    &  $144$     & 12  &  36.6   & 336  &  19  & 37.0  &  383 & 21  &  35.5 &  265 & 17  &  32.8\\
            $\Bpkpi\pi^0$                           & $(-30,~20)$    &  $246$     & 25  &  17.8   & 607  &  40  & 18.3  &  845  & 37  &  17.3 &  583 & 39  &  15.7 \\
            $\Bpks\pi^0$                            & $(-30,~20)$    &  $76$       & 12  &  18.7   & 212  &  20  & 18.8  &  269  & 22  &  18.3  &  185 & 20  &  10.9\\
            $\bar{\Lambda}\pi^-\pi^0$        & $(-30,~20)$    &  $239$     & 18  &  18.1   & 529  &  30  & 17.8  &  723  & 33  &  18.0  &  513 & 28  &  16.9 \\
            $\Bpks\pi^+\pi^-$                     & $(-20,~20)$    &  $89$       & 13  &  20.1   & 192  &  20  & 19.3  &  292  & 24 &  19.7 &  202 & 20  &  17.6\\
            $\bar{\Lambda}\pi^-\pi^+\pi^-$ & $(-20,~20)$    &  $152$     & 14  &  13.6   & 326  &  21  & 14.7  &  453 & 25  &  17.5&  348 & 22  &  13.2\\
            $\bar{\Sigma}^0\pi^-$              & $(-20,~20)$    &  $58$       & 8  &  22.5     & 152  &  13  & 22.0  &  196 & 15  &  20.9 &  145 & 13  &  20.5\\
            $\bar{\Sigma}^-\pi^+\pi^-$       & $(-30,~20)$    &  $136$     & 16  &  19.9   & 362  &  32  & 19.6  &  500  & 32  &  19.9 &  373 & 26  &  17.5\\
      \hline\hline
    \end{tabular}
    \label{tab:yield-st-460}
  \end{center}
\end{table*}

 \begin{figure*}[!htp]
    \begin{center}
        \subfigure[]{\includegraphics[width=0.4\textwidth]{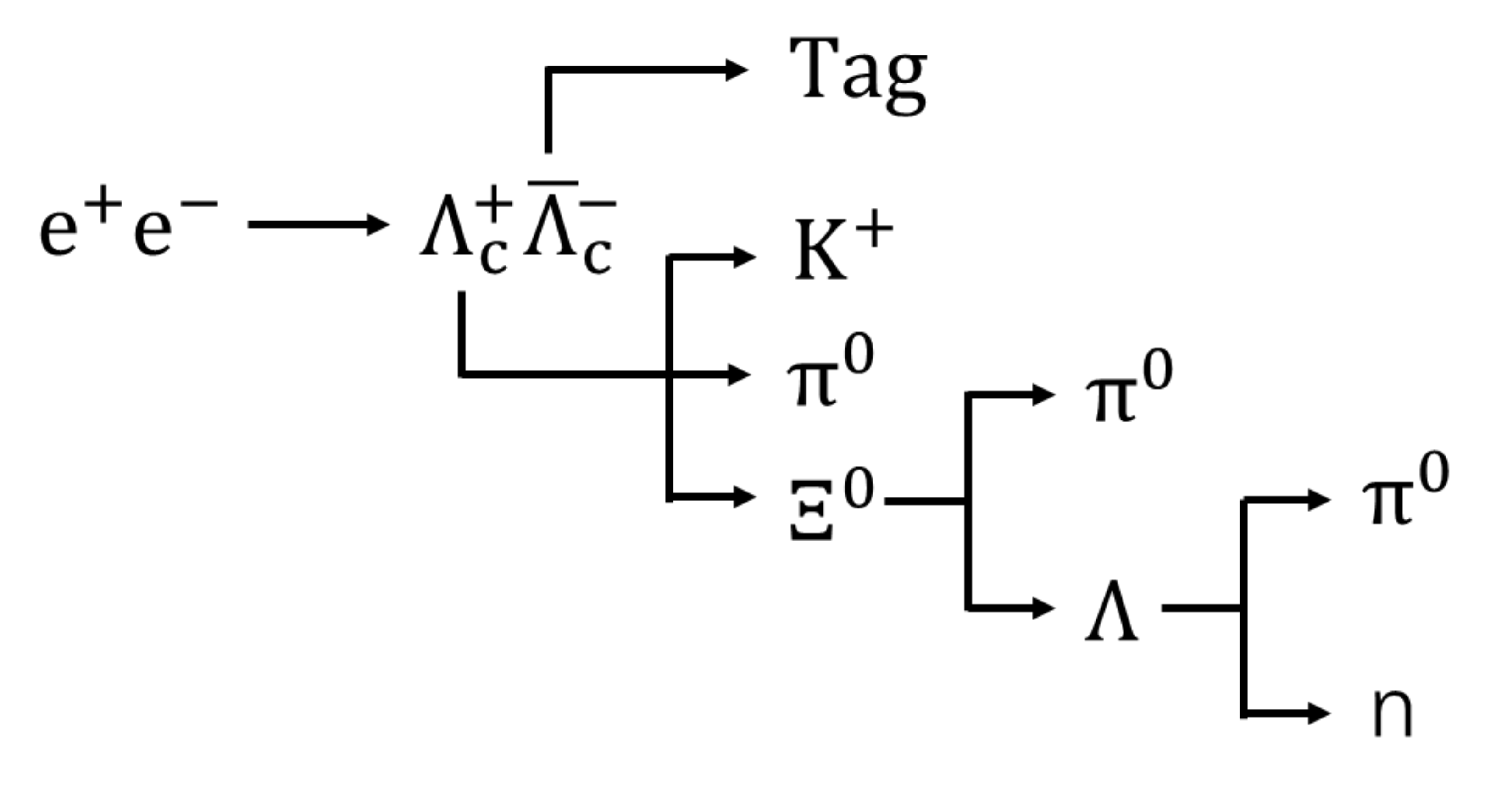}}
          \hspace{3em}%
        \subfigure[]{\includegraphics[width=0.4\textwidth]{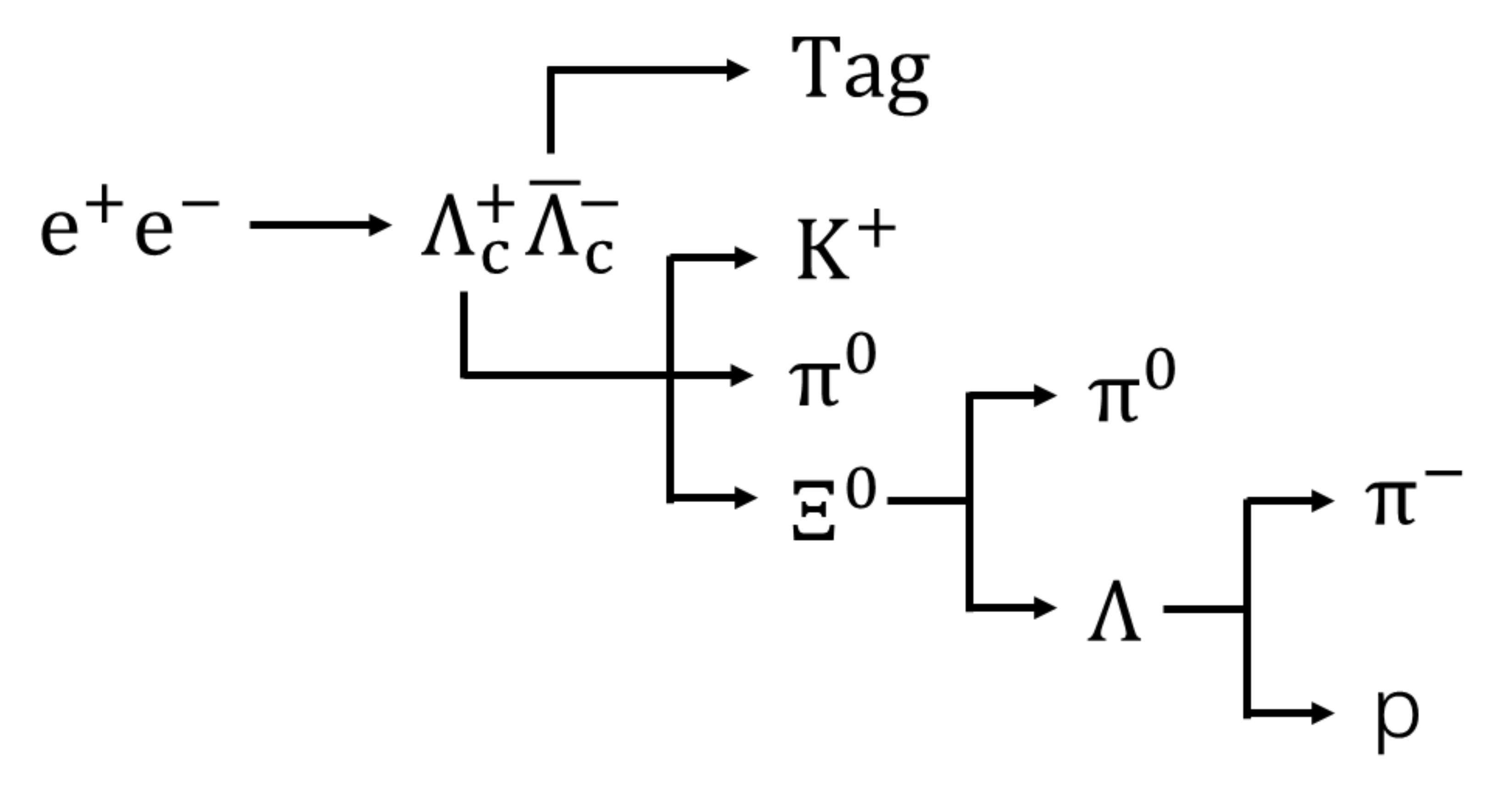}}
    \end{center}
    \caption{
      The schematic diagrams of the two analysis strategies, which are
      denoted (a) Cat-1 and (b) Cat-2  in the
      text.}\label{fig:schematic_diagrams} \end{figure*}

\begin{table*}[htbp]
  \begin{center}
  \caption{The DT detection efficiency(\%) of \LamCXistarK{} in Cat-1/Cat-2 
   for each tag mode and each energy point.}\label{case1-xistar0k}
        \begin{tabular}{ l r @{/} p{0.4 cm}   r @{/} p{0.4 cm}   r @{/} p{0.4 cm}   r @{/} p{0.4 cm}   r @{/} p{0.4 cm}  r @{/} p{0.4 cm}   r @{/} p{0.4 cm}  r @{/} p{0.4 cm}  r @{/} p{0.4 cm}  r @{/} p{0.4 cm}  r @{/} p{0.4 cm} }
      \hline
      \hline
& \multicolumn{2}{c}{4.600} & \multicolumn{2}{c}{4.612} & \multicolumn{2}{c}{4.620} & \multicolumn{2}{c}{4.640} & \multicolumn{2}{c}{4.660} & \multicolumn{2}{c}{4.680} & \multicolumn{2}{c}{4.700} & \multicolumn{2}{c}{4.740} & \multicolumn{2}{c}{4.750} & \multicolumn{2}{c}{4.780} & \multicolumn{2}{c}{4.840} \\
      \hline
         $\pkpi$                                       &  20.3&8.3    &  19.3&7.4    &  19.0&6.7   &  18.6&6.7    & 18.1&6.5     & 17.8&6.5    & 17.4&6.5    &  15.7&6.5   & 15.4&6.4   & 14.9&6.5   & 13.8&6.1    \\
         $\pks$                                        &  22.1&9.0    &  20.4&7.6    &  18.7&8.0   & 18.9&8.1     & 18.4&6.7     & 17.4&8.2    & 17.5&7.5    & 17.3&7.1   & 16.8&7.1   & 17.0&6.6   & 14.0&6.6   \\
         $\Lambda\pi^+$                 &  16.9&7.9    & 15.6&6.9     &  14.5&6.2   &   14.5&6.0   & 13.3&6.1     & 13.0&5.6     & 13.6&4.9   &  13.9&5.3   & 13.4&5.4   & 12.1&5.1   & 12.1&4.6  \\   
         $\pkpi\pi^0$                               &  5.3&1.6      &4.8&1.4        & 4.6&1.5      & 4.5&1.3       & 4.4&1.3       & 4.3&1.3       & 4.1&1.3     &  3.6&1.4     & 3.5&1.4     & 3.5&1.5     & 3.3&1.4 \\
         $\pks\pi^0$                                &  8.0&2.9      & 7.1&2.3       &  7.1&2.0    &   6.8&2.1      & 6.5&2.2      & 6.6&2.5       & 6.0&2.1    &  5.6&2.1     & 5.3&2.2     & 5.1&2.1    & 4.8&2.0 \\
         $\Lambda\pi^+\pi^0$         &  6.7&2.0      &  6.0&1.7     & 5.8&1.8    &   5.7&1.8     & 5.4&1.6       & 5.2&1.7       & 5.1&1.7     &  4.5&1.8     & 4.7&1.7     & 4.4&1.7     & 4.1&1.8  \\
         $\pks\pi^+\pi^-$                         &  9.0&2.9     & 8.7&2.4        & 8.1&2.0     &   7.9&2.2      & 7.9&2.2       & 7.4&2.2      & 7.4&2.6     &  6.7&2.5      & 7.0&2.7    & 6.7&2.8   & 6.1&2.5 \\
         $\Lambda\pi^+\pi^+\pi^-$  &  6.2&1.7     & 5.7&1.3        &  5.4&1.6    &  5.3&1.3       & 5.1&1.4       & 5.3&1.4      & 5.2&1.4     &  4.7&1.6      & 4.5&1.7     & 4.2&1.6     & 4.4&1.8  \\
         $\Sigma^0\pi^+$                       &  8.8&3.7     &7.8&3.0       & 7.8&2.4     &   7.3&2.8      & 6.4&2.4       & 6.6&3.4     & 6.9&2.6     &  5.8&2.5      & 6.0&2.3     & 5.4&2.4    & 5.0&2.3 \\
         $\Sigma^+\pi^+\pi^-$                &  3.2&2.5     & 3.2&2.3       &  3.0&2.3     &   3.0&2.4      & 2.6&2.0       & 2.8&1.9      & 2.7&2.2     &  2.3&1.0      & 2.4&1.0     & 2.2&1.1     & 2.0&0.8 \\
      \hline\hline
    \end{tabular}
  \end{center}
\end{table*}

\begin{table*}[htbp]
  \begin{center}
  \caption{The DT detection efficiency(\%) of \LamCXiKPi{} in Cat-1/Cat-2 
  for each tag mode and each energy point.}\label{case1-xi0kpi0}
      \begin{tabular}{ l r @{/} p{0.56 cm}   r @{/} p{0.56 cm}   r @{/} p{0.56 cm}   r @{/} p{0.56 cm}   r @{/} p{0.56 cm}  r @{/} p{0.56 cm}   r @{/} p{0.56 cm}  r @{/} p{0.56 cm}  r @{/} p{0.56 cm}  r @{/} p{0.56 cm}  r @{/} p{0.56 cm} }
      \hline
      \hline
          & \multicolumn{2}{c}{4.600} & \multicolumn{2}{c}{4.612} & \multicolumn{2}{c}{4.620} & \multicolumn{2}{c}{4.640} & \multicolumn{2}{c}{4.660} & \multicolumn{2}{c}{4.680} & \multicolumn{2}{c}{4.700} & \multicolumn{2}{c}{4.740} & \multicolumn{2}{c}{4.750} & \multicolumn{2}{c}{4.780} & \multicolumn{2}{c}{4.840} \\
      \hline
         $\pkpi$                                       &  7.7& 3.3 &7.4    &  2.8&7.3   &  2.8   &7.2   & 2.7&7.0   & 2.6&7.1   & 2.7&6.8    &  2.8&6.3  & 2.6&6.3 & 2.8&5.9  & 2.8&5.7 & 2.4\\
         $\pks$                                        &  8.6& 3.7 &8.5    &  3.2&7.6    & 3.1   &7.7   & 2.8&7.7   & 2.9&7.7   & 3.2&7.3    &  2.5&6.5  & 3.3&6.7 & 2.9&6.6  & 3.2&6.3  &2.4\\
         $\Lambda\pi^+$                 &  6.3& 2.4 &6.0    &  2.0&6.1    &   2.2  &6.1   & 2.1&5.1   & 2.1&5.5  & 1.9&5.9    &  2.4&5.3  & 2.2&5.5 & 2.3&5.2  & 2.5&4.7  &2.1\\   
         $\pkpi\pi^0$                               &  1.9&0.8 & 2.0    &  0.7&2.0    &   0.7  &1.9   & 0.6&2.0  & 0.6&1.8   & 0.7&1.8    &  0.6&1.5  & 0.6&1.4 & 0.6&1.5  & 0.6&1.3  &0.6\\
         $\pks\pi^0$                                &  2.9&1.1 &2.5     &  1.1&2.5    &   1.0 &2.3   & 1.0&2.5   & 1.1&2.6   & 0.9&2.4   &  1.0&2.1  & 0.7&2.1 & 0.8&2.0  & 0.9&2.2   &0.8 \\
         $\Lambda\pi^+\pi^0$         &  2.5& 0.9 &2.2    & 0.8&2.1    &   0.8  &2.1   & 0.8&2.1   & 0.7&2.0   & 0.8&2.0    &  0.8&1.8  & 0.7&1.7 & 0.7&1.9  & 0.8&1.8   &0.7\\
         $\pks\pi^+\pi^-$                         &  3.2&1.4 &2.7     & 1.0&3.0    &   1.0  &3.1   & 1.0&2.6   & 1.1&2.7   & 0.9&3.0    &  1.1&2.4  & 0.9&2.8 & 1.0&2.7  & 1.0&2.5   &1.1\\
         $\Lambda\pi^+\pi^+\pi^-$  &  2.2&0.8 &2.0     &  0.6&1.8   &   0.6. &2.0   & 0.6&1.9   & 0.6&1.9   & 0.6&1.9    &  0.7&1.8  & 0.7&1.6 & 0.6&1.6  & 0.6&1.8   &0.6\\
         $\Sigma^0\pi^+$                       &  3.2&1.3 &2.8     &  1.3&2.9    &   1.1 &2.5   & 1.3&2.8   & 1.3&2.9   & 1.1&2.5    &  1.0&2.7 & 1.1&2.4 & 1.2&2.2  & 1.0&2.2    &1.0\\
         $\Sigma^+\pi^+\pi^-$                &  1.3&0.7 &1.2     &  0.4&1.1    &   0.4. &1.1   & 0.5&1.1   & 0.4&1.0   & 0.5&1.0    &  0.4&0.8  & 0.5&0.8 & 0.4&0.8  & 0.4&0.9  &0.3\\
      \hline\hline
    \end{tabular}
  \end{center}
\end{table*}

\begin{table*}[htbp]
  \begin{center}
  \caption{The DT detection efficiency(\%) of \LamCNKPi{} in Cat-1 for each tag mode and each energy point.}\label{case1-nkpi0}
    \begin{tabular}{ l c c c c c c c c c c c}
      \hline
      \hline
          & 4.600 & 4.612 & 4.620 & 4.640 & 4.660 & 4.680 & 4.700 & 4.740 & 4.750 & 4.780 & 4.840 \\
      \hline
         $\pkpi$                                       &  13.4    & 12.9   &  12.2   &  12.0   & 11.7   & 11.4   & 11.2    &  10.4  & 10.2& 9.9  & 9.2    \\
         $\pks$                                        &  14.3    & 13.8   &  13.3   & 12.9  & 12.1 & 12.5      & 11.5 &  11.2  & 11.6 & 11.1  & 9.9   \\
         $\Lambda\pi^+$                 &  11.3    & 10.6    &  9.7  &   10.0   & 9.3   & 9.4   & 9.5    &  9.3  & 8.2& 7.9 & 8.1  \\   
         $\pkpi\pi^0$                               &  3.7      & 3.6      &  3.5    &   3.3   & 3.3   & 3.2     & 3.2    &  2.7  & 2.7 & 2.8  & 2.5 \\
         $\pks\pi^0$                                &  4.8     & 4.5    &  4.3    &   4.5   & 4.2  & 4.0     & 3.8   &  3.1  & 3.4 & 3.4  & 3.0 \\
         $\Lambda\pi^+\pi^0$         &  4.2     & 4.0      & 3.6     &   3.7   & 3.5   & 3.4     & 3.4    &  3.2 & 3.1 & 3.0  & 2.7  \\
         $\pks\pi^+\pi^-$                         &  5.8    & 5.2       & 4.8     &   4.9   & 4.6   & 4.5      & 4.8    &  4.3  & 3.8 & 4.3  & 4.2 \\
         $\Lambda\pi^+\pi^+\pi^-$  &  3.9    & 3.3      &  3.5    &   3.2   & 3.4  & 3.4   & 2.9    &  2.7  & 2.8 & 2.7 & 2.8  \\
         $\Sigma^0\pi^+$                       &  6.7     & 5.8      &  6.2   &   5.5   & 5.2   & 5.5     & 6.0   &  4.3  & 5.0 & 4.0  & 4.1 \\
         $\Sigma^+\pi^+\pi^-$                &  2.4     & 2.3       &  2.1    &   2.3  & 2.2   & 2.2     & 2.0    &  1.8  & 1.9 & 1.9  & 1.6 \\
      \hline\hline
    \end{tabular}
  \end{center}
\end{table*}

\begin{table*}[htbp]
  \begin{center}
  \caption{The DT detection efficiency(\%) of \LamCSigKPi{} in Cat-1/Cat-2 
  for each tag mode and each energy point.}\label{case1-sigma0kpi0}
     \begin{tabular}{ l r @{/} p{0.52 cm}   r @{/} p{0.52 cm}   r @{/} p{0.52 cm}   r @{/} p{0.52 cm}   r @{/} p{0.52 cm}  r @{/} p{0.52 cm}   r @{/} p{0.52 cm}  r @{/} p{0.52 cm}  r @{/} p{0.52 cm}  r @{/} p{0.52 cm}  r @{/} p{0.52 cm} }
      \hline
      \hline
& \multicolumn{2}{c}{4.600} & \multicolumn{2}{c}{4.612} & \multicolumn{2}{c}{4.620} & \multicolumn{2}{c}{4.640} & \multicolumn{2}{c}{4.660} & \multicolumn{2}{c}{4.680} & \multicolumn{2}{c}{4.700} & \multicolumn{2}{c}{4.740} & \multicolumn{2}{c}{4.750} & \multicolumn{2}{c}{4.780} & \multicolumn{2}{c}{4.840} \\
      \hline
         $\pkpi$                                       &  10.2&5.7    & 9.7&5.3    &  9.0&4.9   &  9.3&4.9      & 9.2&4.9   & 8.7&4.8   & 8.6&4.7    &  8.3&4.5  & 8.1&4.6 & 7.7&4.5  & 7.5&4.2    \\
         $\pks$                                        &  11.4&6.1    & 10.0&5.5   &  9.7&5.3   & 9.9&5.6      & 10.0&5.5 & 9.0&5.0   & 9.4&4.6    &  8.8&5.5  & 8.6&5.4 & 8.5&4.9  & 8.2&4.8   \\
         $\Lambda\pi^+$                 &  8.5&4.8      & 8.5&3.6     &  7.3&3.9   & 7.4&4.3     & 7.7&3.9   & 7.0&4.0   & 6.8&3.3    &  6.9&3.9  & 6.9&4.2 & 6.2&3.7  & 6.0&3.5  \\   
         $\pkpi\pi^0$                               &  2.7&1.2      & 2.4&1.1      &  2.5&1.1  &   2.4&1.1   & 2.3&1.1   &2.2&0.9      & 2.4&1.0    &  2.0&0.9  & 2.0&1.0 & 1.9&1.0  & 1.9&1.0 \\
         $\pks\pi^0$                                &  4.1&1.7      & 3.3&2.0      &  3.4&1.7  &   3.2&1.8   & 3.0&1.7   & 3.4&1.6      & 3.1&1.8    &  3.3&1.5  & 2.7&1.4 & 2.7&1.7 & 2.4&1.3 \\
         $\Lambda\pi^+\pi^0$         &  3.1&1.7      & 2.9&1.3      & 2.7&1.4   &   2.8&1.4   & 2.8&1.2  & 2.8&1.3      & 2.5&1.2    &  2.3&1.3  & 2.3&1.3 & 2.4&1.2  & 2.1&1.2  \\
         $\pks\pi^+\pi^-$                         &  4.5&2.1     & 4.1&1.8       & 3.6&1.9   &   3.8&1.9   & 3.8&1.5   & 3.4&1.7      & 3.8&1.7    &  3.7&1.6  & 3.0&1.5 & 3.5&1.6  & 3.4&1.8 \\
         $\Lambda\pi^+\pi^+\pi^-$  &  3.1&1.4    & 2.6&1.2       &  2.5&1.1   &   2.7&1.2    & 2.7&1.3   & 2.5&1.1      & 2.5&1.2    &  2.2&1.1  & 2.2&1.2 & 2.2&1.2  & 2.2&1.2  \\
         $\Sigma^0\pi^+$                       &  4.4&2.1     & 4.6&2.1       &  3.7&2.2   &   3.6&1.9   & 3.9&1.6   & 3.8&1.8     & 3.6&1.9    &  3.4&2.0  & 3.5&1.7 & 3.2&1.7  & 3.1&1.7 \\
         $\Sigma^+\pi^+\pi^-$                &  1.8&1.0     & 1.6&0.8       &  1.5&0.8   &   1.5&0.8   & 1.5&0.8   & 1.4&0.9     & 1.6&0.7    &  1.2&0.7  & 1.2&0.7 & 1.3&0.8  & 1.2&0.8  \\
      \hline\hline
    \end{tabular}
  \end{center}
\end{table*}

\begin{table*}[htbp]
  \begin{center}
  \caption{The DT detection efficiency(\%) of \LamCLamKPi{} in Cat-1/Cat-2
  for each tag mode and each energy point.}\label{case1-lambdakpi0}
    \begin{tabular}{ l r @{/} p{0.5 cm}   r @{/} p{0.5 cm}   r @{/} p{0.5 cm}   r @{/} p{0.5 cm}   r @{/} p{0.5 cm}  r @{/} p{0.5 cm}   r @{/} p{0.5 cm}  r @{/} p{0.5 cm}  r @{/} p{0.5 cm}  r @{/} p{0.5 cm}  r @{/} p{0.5 cm} }
      \hline
      \hline
& \multicolumn{2}{c}{4.600} & \multicolumn{2}{c}{4.612} & \multicolumn{2}{c}{4.620} & \multicolumn{2}{c}{4.640} & \multicolumn{2}{c}{4.660} & \multicolumn{2}{c}{4.680} & \multicolumn{2}{c}{4.700} & \multicolumn{2}{c}{4.740} & \multicolumn{2}{c}{4.750} & \multicolumn{2}{c}{4.780} & \multicolumn{2}{c}{4.840} \\
      \hline
         $\pkpi$                                       &  11.0&6.4    & 10.5&6.0   &  10.0&5.6   &  10.0&5.4   & 10.1&5.4   & 9.9&5.5   & 9.5&5.4    &  8.7   &5.0  & 8.6&5.1& 8.2&5.0  & 7.9&4.6    \\
         $\pks$                                        &  11.9&7.2   & 10.9&6.7   &  10.9&6.2   & 11.1&6.1    & 10.5&6.0  & 9.9&5.6    & 9.6&5.6    &  9.6    &5.8  & 9.4&5.9 & 9.3&5.9  & 8.5&5.3   \\
         $\Lambda\pi^+$                 &  9.8&4.9      & 9.0&4.8     &  7.9&4.5     &   8.6&5.0   & 8.0&4.5  & 7.1&4.6       & 7.8&4.3    &  7.6    &4.3  & 7.3&4.8& 8.2&4.3 & 7.4&3.7  \\   
         $\pkpi\pi^0$                               &  3.0&1.4      & 2.9&1.3      &  2.8&1.3    &   2.8&1.2   & 2.6&1.2   & 2.6&1.2      & 2.7&1.1     &  2.3   &1.2 & 2.2&1.2 & 2.2&1.1  & 2.1&1.1 \\
         $\pks\pi^0$                                &  4.6&2.3      & 3.6&1.9      &  3.9&1.7     &   3.5&1.8   & 3.2&1.8   & 3.4&1.9      & 3.5&1.5    &  2.7   &1.7  & 3.2&1.6 & 2.9&1.6  & 2.5&1.6 \\
         $\Lambda\pi^+\pi^0$         &  3.5&1.9      & 3.2&1.7      & 3.1&1.6     &   3.1&1.6    & 3.1&1.5   & 3.0&1.5      & 2.7&1.5     &  2.6   &1.4 & 2.4&1.3 & 2.4&1.3  & 2.3&1.3  \\
         $\pks\pi^+\pi^-$                         &  4.7&2.4      & 4.1&2.1       & 4.0&1.9     &   4.4&1.9   & 4.2&1.8   & 3.8&2.0      & 4.1&1.8     &  3.8   &2.0  & 3.4&2.0 & 4.1&2.1  & 3.5&1.9 \\
         $\Lambda\pi^+\pi^+\pi^-$  &  3.0&1.7      & 2.7&1.3       &  2.7&1.4   &   2.7&1.4   & 2.7&1.3   & 2.8&1.4      & 2.9&1.4     &  2.4   &1.4  & 2.8&1.3 & 2.5&1.3  & 2.3&1.4  \\
         $\Sigma^0\pi^+$                       &  5.1&2.8     & 4.5&2.4       &  4.4&2.7    &   4.6&2.4   & 4.2&2.4   & 4.4&2.1     & 4.0&2.2      &  3.8   &2.6  & 3.4&2.5 & 3.3&2.2  & 3.0&2.1 \\
         $\Sigma^+\pi^+\pi^-$                &  2.1&1.4      & 1.9&1.1       &  1.8&1.1    &   1.6&1.0   & 1.8&1.1   & 1.6&1.0    & 1.5&0.9      &  1.6   &0.8   & 1.5&0.9 & 1.3&0.8  & 1.3&0.8  \\
      \hline\hline
    \end{tabular}
  \end{center}
\end{table*}

After the ST selection, we want to determine the number of
$\LamCXiKPi{}$, $nK^{+}\pi^{0}$, $\Sigma^{0}K^{+}\pi^{0}$ and $\Lambda K^{+}\pi^{0}$ candidates on the opposite side of
the ST \LamCBar. The $K^{+}$ and $\pi^{0}$ are reconstructed, as
described above, from the remaining showers and tracks.  Using total
four-momentum conservation, the $\Xi^{0}$ signal is obtained from the
distribution of recoil mass,

\begin{equation} \label{eq:mrecKpi}
\begin{aligned}
  M^{2}_{\mathrm{recoil}}(\Lambda _c^{\mathrm{ST}} {K^ + }{\pi ^ 0 })  = (E_{\mathrm{beam}}- E_{K^ +} - E_{\pi^0})^2/c^4 \\ - | \rho\cdot\vec{p}_{0} -\vec{p}_{K^ +} - \vec{p}_{\pi^0} |^2/c^2, 
\end{aligned}
\end{equation}

\noindent where $E_{K^ +}$, $E_{\pi^0}$, $\vec{p}_{K^ +}$ and
$\vec{p}_{\pi^0}$ are the energies and momenta of ${K^ +}$ and $\pi^0$
candidates, respectively, $\rho = \sqrt{E_{\mathrm{beam}}^2/c^2 -
m_{\LamC}^2 c^2}$ is the magnitude of the $\Lambda^{+}_{c}$ momentum,
constrained by the beam energy, and $\vec{p}_{0} = -\vec{p}_{\LamCB}
/|\vec{p}_{\LamCB}|$ is its direction.
Similarly, the yield of the two-body decay $\LamCXistarK$ is obtained
using the recoil mass.
\begin{equation} \label{eq:mrecKA} M^{2}_{\mathrm{recoil}}(\Lambda
_c^{\mathrm{ST}} {K^ + }) = (E_{\mathrm{beam}}- E_{K^ +})^2/c^4 - |
\rho\cdot\vec{p}_{0} -\vec{p}_{K^ +} |^2/c^2.  \end{equation}

The $\Xi(1530)^0$, $\Xi^0$ and $\Sigma^0$ predominant decay modes are $\Xi(1530)^{0} \to \Xi \pi$, $\XiLamPi{}$ and $\Sigma^{0} \to \Lambda \gamma$~\cite{g3}, respectively. The
$\Lambda $ baryon has two main decay modes, which are studied in this
analysis: the neutral decay mode $\Lambda\to n\pi^{0}$, which will be
referred to as Cat-1, and the charged decay mode $\Lambda\to p\pi^{-}$,
which will be referred to as Cat-2.  The Cat-1 and Cat-2 decays are shown schematically in Fig.~\ref{fig:schematic_diagrams}.
Four variables $\MrecPKPI$, $\MrecPKPIB$, $\MrecPK$ and $\MrecPKB$ are
defined by Eqs.~\ref{eq:mrecKpi} and~\ref{eq:mrecKA}.
For Cat-1, a single tight track is reconstructed and identified as a
$K^+$ meson.  In Cat-2, three charged tracks are
reconstructed as a $K^+$, $p$, and $\pi^-$. The $K^+$ is a tight track,
while the other two are loose tracks from the $\Lambda\to p\pi^{-}$
decay. The $p$ and $\pi^-$ are constrained to originate from a common
vertex to form the $\Lambda$ candidates by a vertex fit~\cite{vertex}.
The invariant mass of the $\Lambda$ candidate is
required to fall inside the range (1.111, 1.121)~GeV/${c^2}$.
  
The $\pi^0$ candidate is reconstructed via the process
\PiGamGam{}.  If there are multiple $\pi^0$ candidates, the one with
the minimum kinematic-fit $\chi^2$ is assumed to originate from the
$\Lambda^{+}_{c}$ baryon.

To suppress sources of background due to neutral hyperon decays from
the decays $\LamC\to \Lambda K^+$, $\Sigma^0 K^+$ and $\Xi^0 K^+$, the
events falling inside the corresponding intervals of these three
decays of the variable $\MrecPK$ are rejected.  This requirement
suppresses 95\% of these backgrounds.  Furthermore, electron mis-PID
backgrounds are suppressed using PID and EMC information. To suppress
contaminations from long-lived particles in the final state, any
event with additional tracks, either loose or tight, is rejected. The
DT efficiencies of each signal decay for Cat-1 and Cat-2 are
summarized in
Tables~\ref{case1-xistar0k}{\color{blue}{-}}\ref{case1-lambdakpi0}.

\section{Background analysis}

Potential sources of background are classified into$~$two categories:
those from $\ee$ $\to$ $\LCpair$ (denoted as $\LCpair$ background) and
those directly originating from continuum hadron production in $\ee$
annihilation (denoted as $q\bar{q}$ background).

The $\LCpair$ background is investigated for both categories with the
$\LCpair$ inclusive MC samples, after removing the signal processes.
The $\LCpair$ background distributions are the blue-shaded histograms
in Figs.~\ref{fig:fit_Case2} and~\ref{fig:fit_reck}, and they have
been normalized to the same luminosity as data.

For the Cat-1 $q\bar{q}$ background, the data events in the $M_{\rm
  BC}$ sideband region, defined as $M_{\rm BC}\in (2.200,2.265)$
  GeV/$c^2$, are used.  Since $\LCpair$ background is present in
  the sideband region, it is estimated using inclusive simulation
  samples and subtracted from the total sideband yield to
  determine the $q\bar{q}$ background in the sideband.  The $q\bar{q}$
  background in the signal region is extrapolated from the yield in
  the sideband using information from the simulation. In Cat-2, the
  simulated $q\bar q$ samples are used.

There exists a contamination from mismatched $\pi^0$ candidates (denoted as $\pi^0$ mismatch), which consists of three sources: the first is due to the 
selection of an incorrect $\pi^0$ candidate in a signal event with 
multiple $\pi^0$ candidates; the second is where the $\pi^0$ candidate is 
reconstructed using photons that originated from two different $\pi^0$ 
decays; the third is caused by noise in the EMC, which creates a fake 
photon candidate that is used in the $\pi^0$ reconstruction.
It has a wide shape and
will not contribute to the narrow resonances in the distributions of
\MrecPKPI{} and \MrecPKPIB{}.  The
$\pi^0$ mismatch background is estimated using simulation
samples. Events containing showers where the angular separation
between the true and reconstructed $ \pi^{0}$ directions is greater
than $20^{\circ}$ are considered $\pi^0$ mismatch background, and
the background is normalized according to the branching fractions of
$\LamCXistarK$ and $\LamCXiKPi{}$ in the distributions of $\MrecPKPI$
and $\MrecPKPIB$.

\section{Branching fraction measurement}

For the two-body decay $\LamCXistarK$, a simultaneous fit is performed
by combining the Cat-1 and Cat-2 distributions of \MrecPK{} and
$\MrecPKB$. The signal shapes are modeled with the simulation shapes
convolved with a Gaussian function representing the resolution
difference between data and simulation samples.  The background shapes
are described by third-order polynomial functions, and the background
yields are floating. The fitted curves with distributions of $\MrecPK$
and $\MrecPKB$ are shown in Fig.~\ref{fig:fit_Case2}. The signal
yields are $30 \pm 6$ and $20 \pm 5$ events for Cat-1 and Cat-2,
respectively. The DT efficiencies in Cat-1 and Cat-2 are summarized in
Table~\ref{case1-xistar0k}. From Eq.~\ref{eq:br1}, the branching
fraction is determined to be $(5.99\pm1.04)\times 10^{-3}$, where the
uncertainty is statistical. It is consistent with the previous
result~\cite{tobepu}. The significance, at $6.9\sigma$, is
determined by evaluating the difference in likelihood values between fits that include and exclude the signal component, while considering the change in the number of degrees of freedom.

For the three-body decay $\LamCXiKPi{}$, the simultaneous fit shares
the same branching fraction and takes into account the different
detection efficiencies of the individual processes.  The fit
projections of $\MrecPKPI$ and $\MrecPKPIB$ are shown in
Fig.~\ref{fig:fit_reck}, and the total signal yields of \LamCXiKPi{} are
$57 \pm 9$ and $40 \pm 7$ events for Cat-1 and Cat-2, respectively.

The DT efficiencies of Cat-1 and Cat-2 for $\LamCXiKPi{}$ are summarized in
Table~\ref{case1-xi0kpi0}. The simulation samples use a flat phase
space model, but intermediate resonances could affect the momentum and
angular distributions of the final state particles, which would affect DT
efficiencies. Therefore, the simulation samples are reweighted to
match data. The two variables, $\MrecKPI$ and $\MrecPI$, are selected
for the reweighting procedure using a 6$\times$6 binning scheme
with uniformly spaced bins.  The overall efficiency is 3.44\% before the
procedure and 3.61\% after.

Using the results of fits to $\MrecPK$ and $\MrecPKB$ and correcting
for the differences in efficiency, the contributions to the three-body
decays $\LamCXiKPi{}$ are calculated.  After subtracting them, the
yields of the three-body decay \LamCXiKPi{} are $48 \pm 9$ and $31 \pm
7$ for Cat-1 and Cat-2, respectively. Using Eq. \ref{eq:br1} and the
corrected efficiency, the branching fraction is determined to be
$(7.79\pm1.46)\times 10^{-3}$, where the uncertainty is statistical,
and the statistical significance is $8.6\sigma$.

\begin{figure*}[!htp]
    \centering
             \subfigure[]{\includegraphics[width= 0.49\textwidth,height=0.25\textheight]{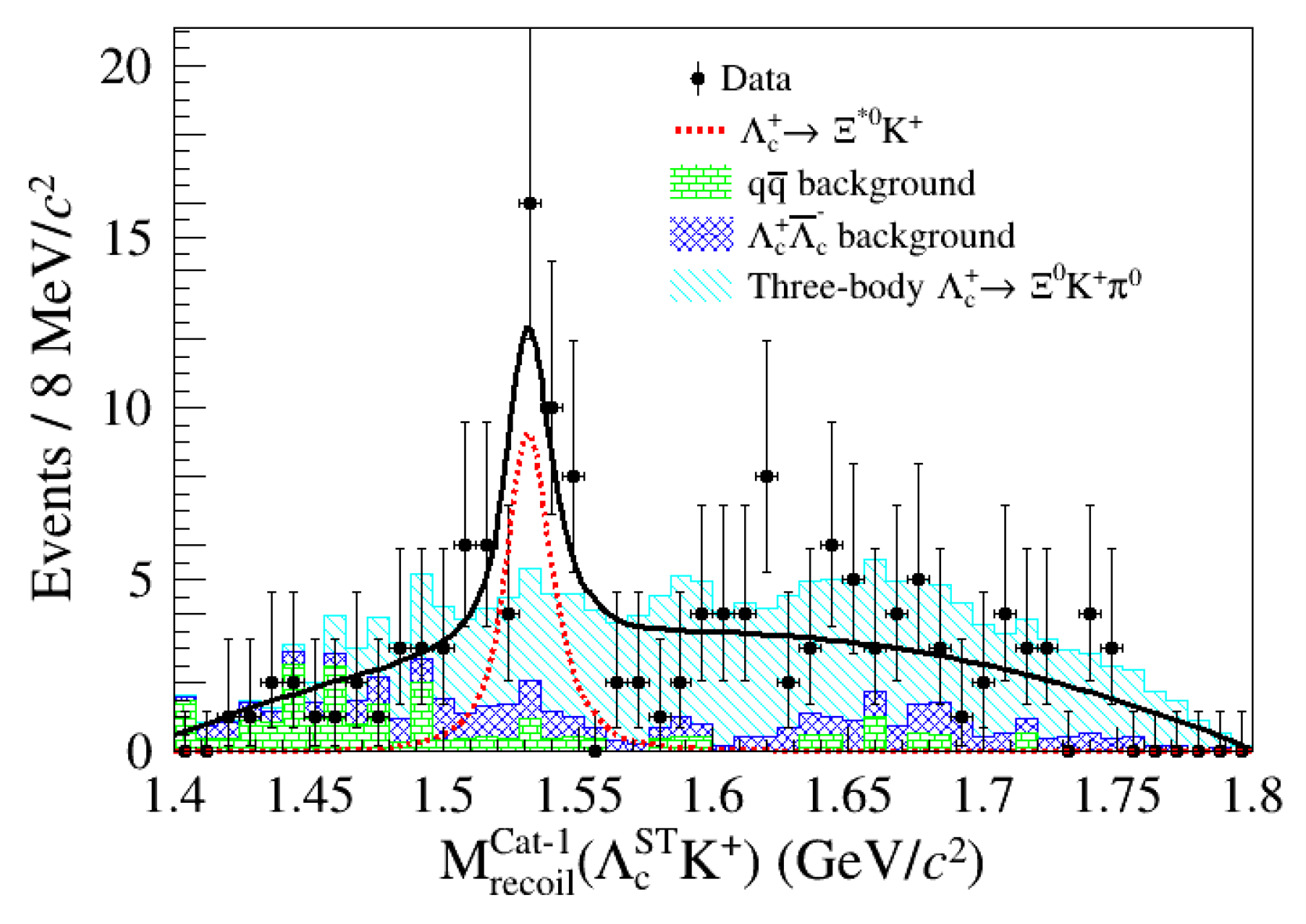}\label{fig:signaly_Case1}}
            \subfigure[]{\includegraphics[width=
            0.49\textwidth,height=0.25\textheight]{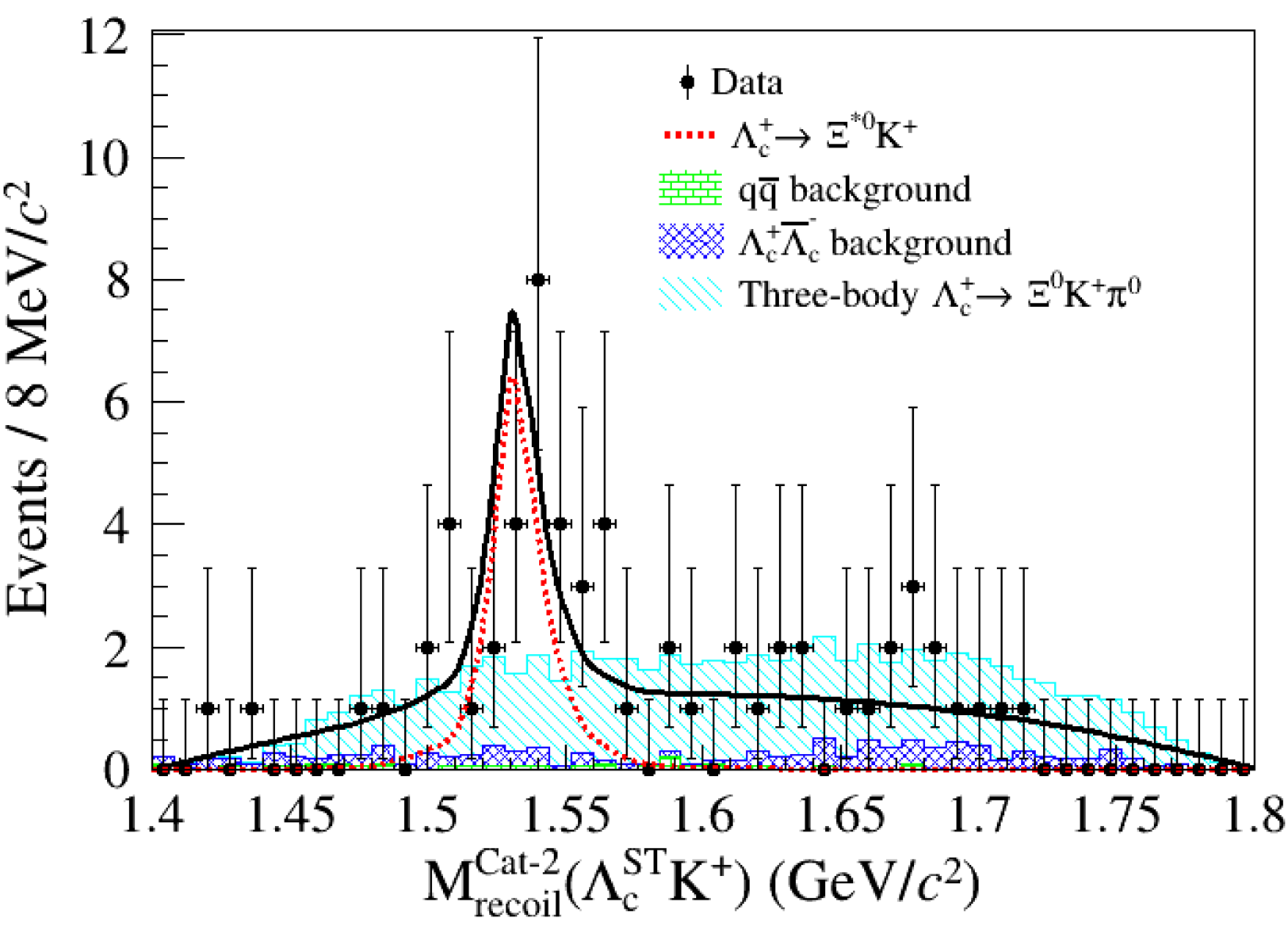}\label{fig:signalx_Case2}}
            \caption{The fit projections of (a) $\MrecPK$ and (b)$\MrecPKB$
            distributions.  The black points with error bars are
            data. The black line is the sum of all the components in
            the fit. The blue-shaded histograms are the
            $\LamC\LamCBar$ backgrounds. The green-shaded histograms are
            from the data sideband (in Cat-1) or hadron sample (in
            Cat-2). }
        \label{fig:fit_Case2}
\end{figure*}
 
 \begin{figure*}[!htp]
    \centering
             \subfigure[]{\includegraphics[width= 0.49\textwidth,height=0.25\textheight]{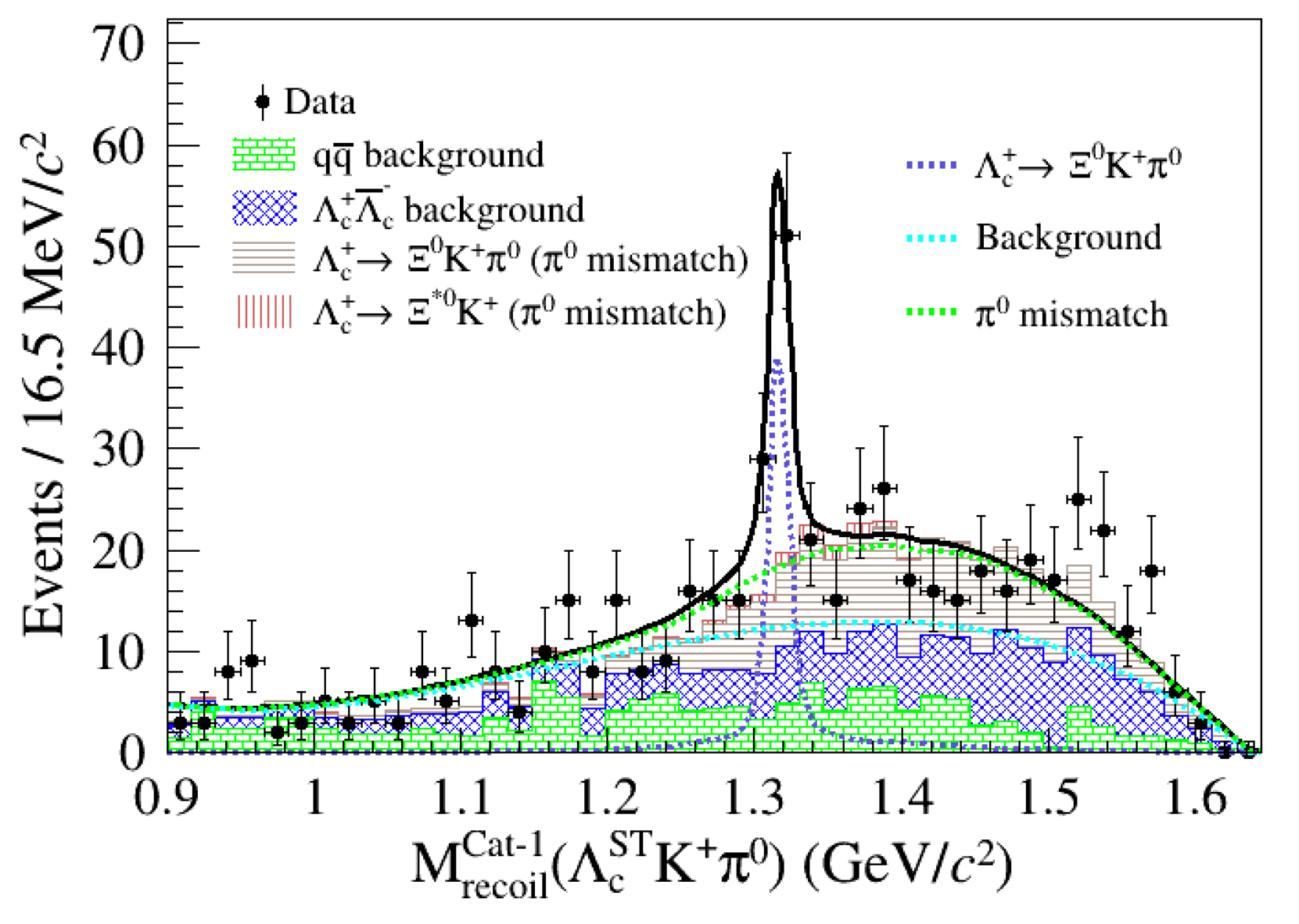}     \label{fig:xi0kpi0_Case1}}       
            \subfigure[]{\includegraphics[width= 0.49\textwidth,height=0.25\textheight]{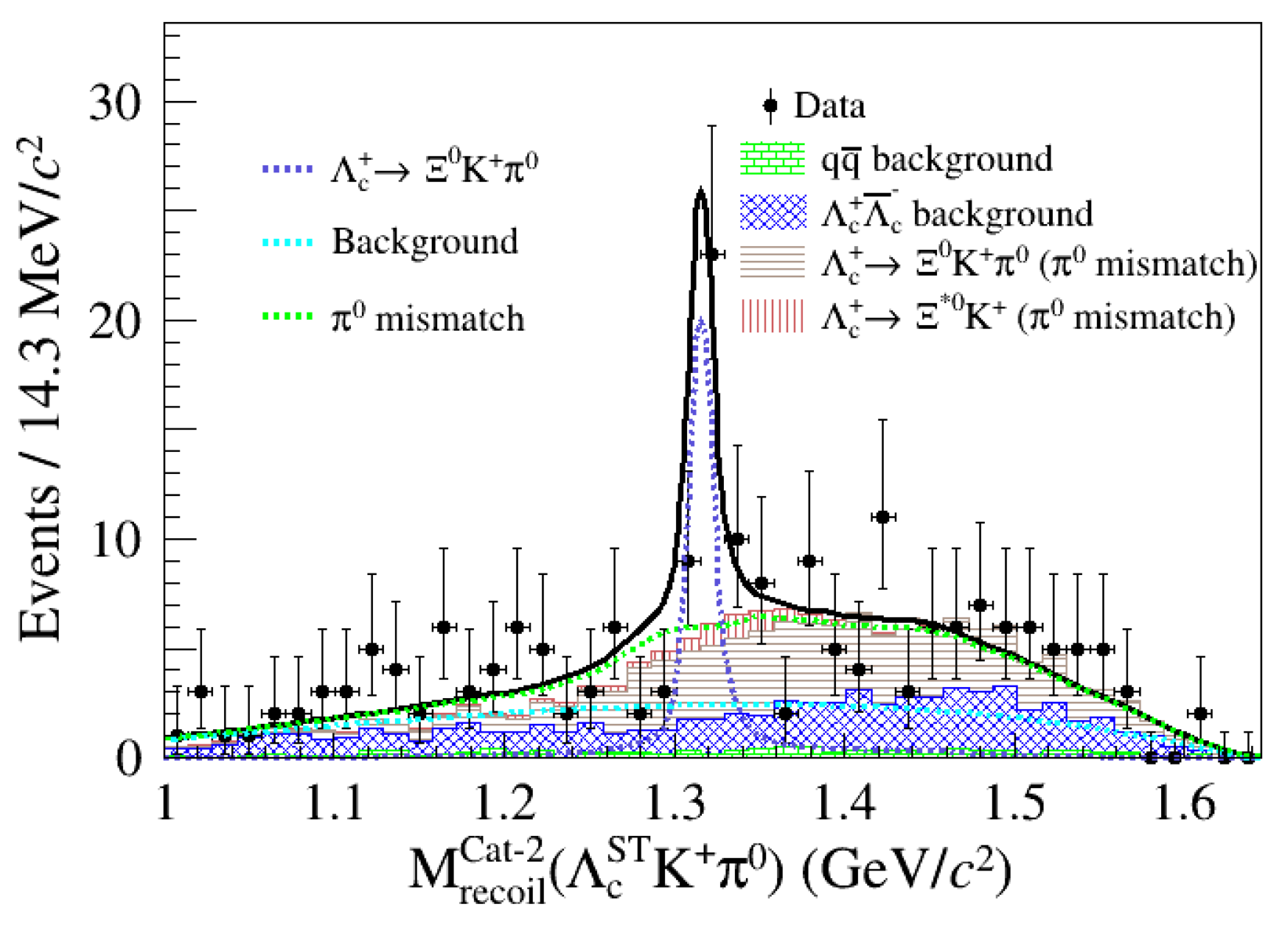} \label{fig:xi0kpi0_Case2}}  
        \caption{Fit projections of (a) $\MrecPKPI$ and (b)
        $\MrecPKPIB$ distributions. The black points with error bars
        are data. The black line is the sum of all the components in
        the fit. The blue-shaded histograms are the
        $\LamC\LamCBar$ background. The green-shaded histograms are from
        the data sideband (in Cat-1) or hadron sample (in Cat-2). The
        purple line represents the signal shape of the three-body
        decay $\LamCXiKPi$. The dashed blue line represents the one-dimensional histogram PDF that describes the background
        distributions of $\MrecPKPIAB$, which are modeled with an polynomial function.}
        \label{fig:fit_reck}
\end{figure*}

 \begin{figure*}[!htp]
 \centering
               \subfigure[]{\includegraphics[width= 0.319\textwidth,height=0.169\textheight]{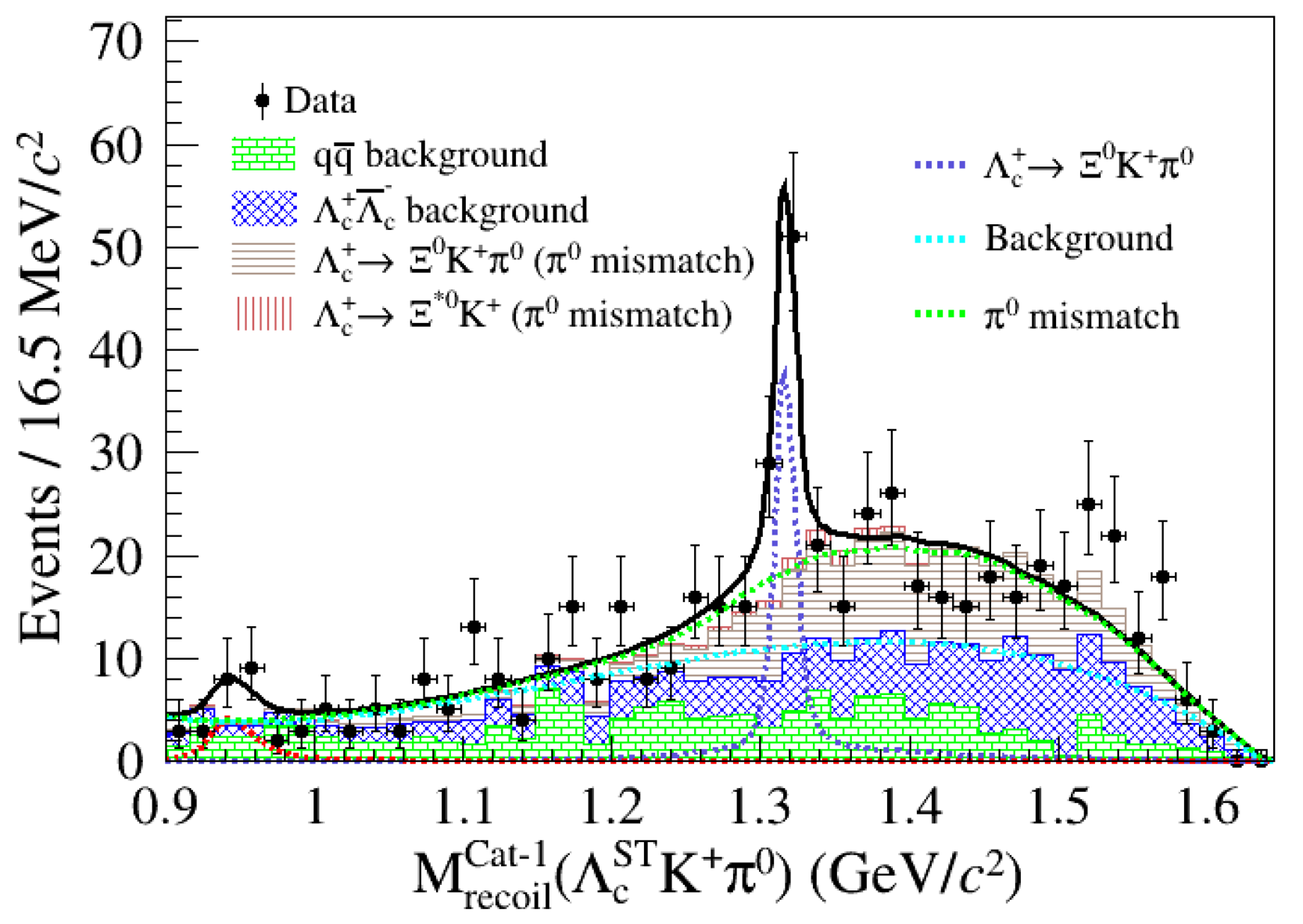}     \label{fig:nkpi0_cat1}}       
            \subfigure[]{\includegraphics[width= 0.319\textwidth,height=0.169\textheight]{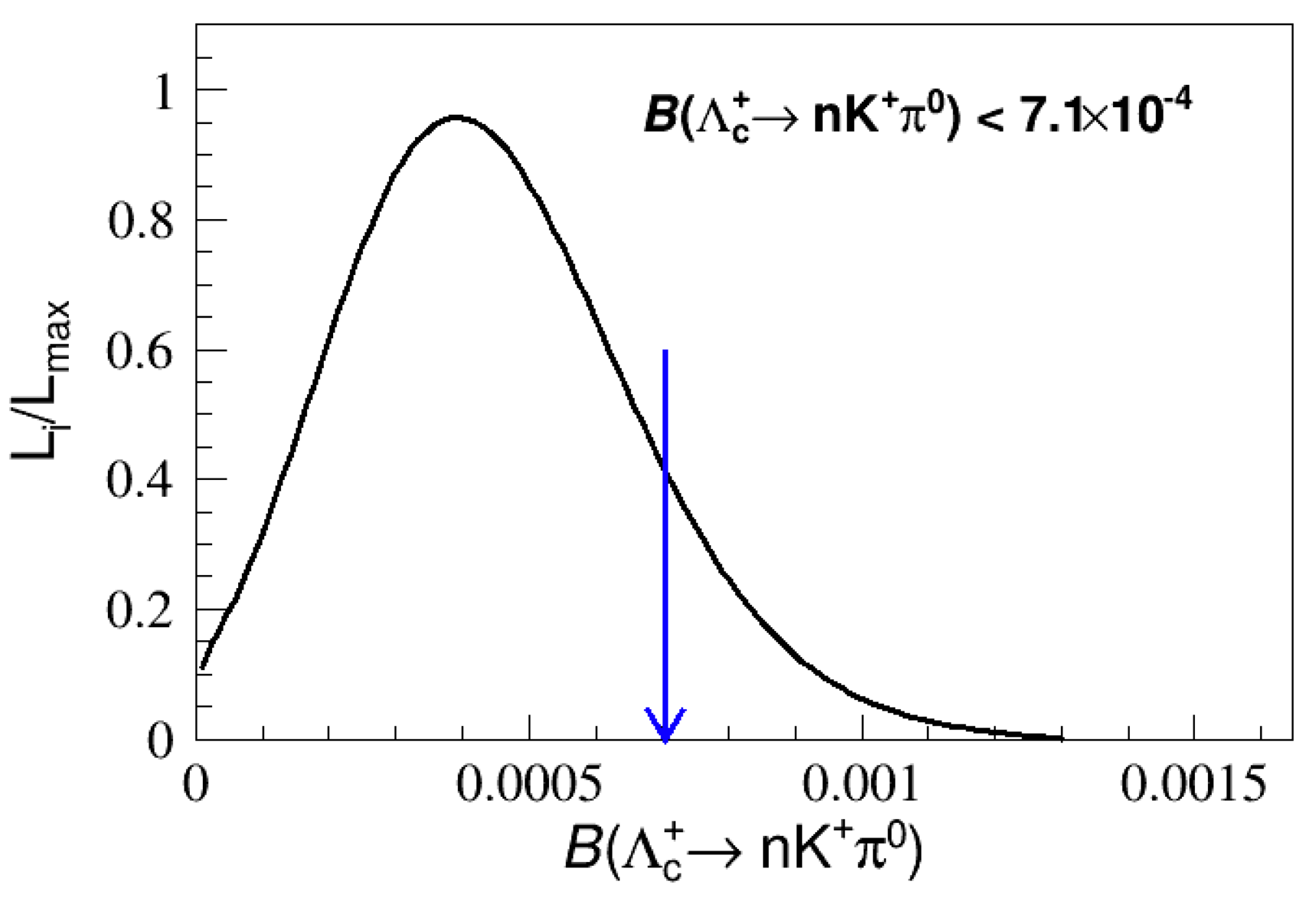} \label{fig:noComfit_uplimit_3}}              
    \caption{Fit projections on (a) $\MrecPKPI$ distribution and (b)
    Likelihood distribution over the branching fractions of $\LamC \to
    n K^{+}\pi^{0}$.
    }\label{fig:nkpi0}
\end{figure*}

 \begin{figure*}[!htp]
 \centering
 \subfigure[]{\includegraphics[width=0.319\textwidth,height=0.226\textwidth]{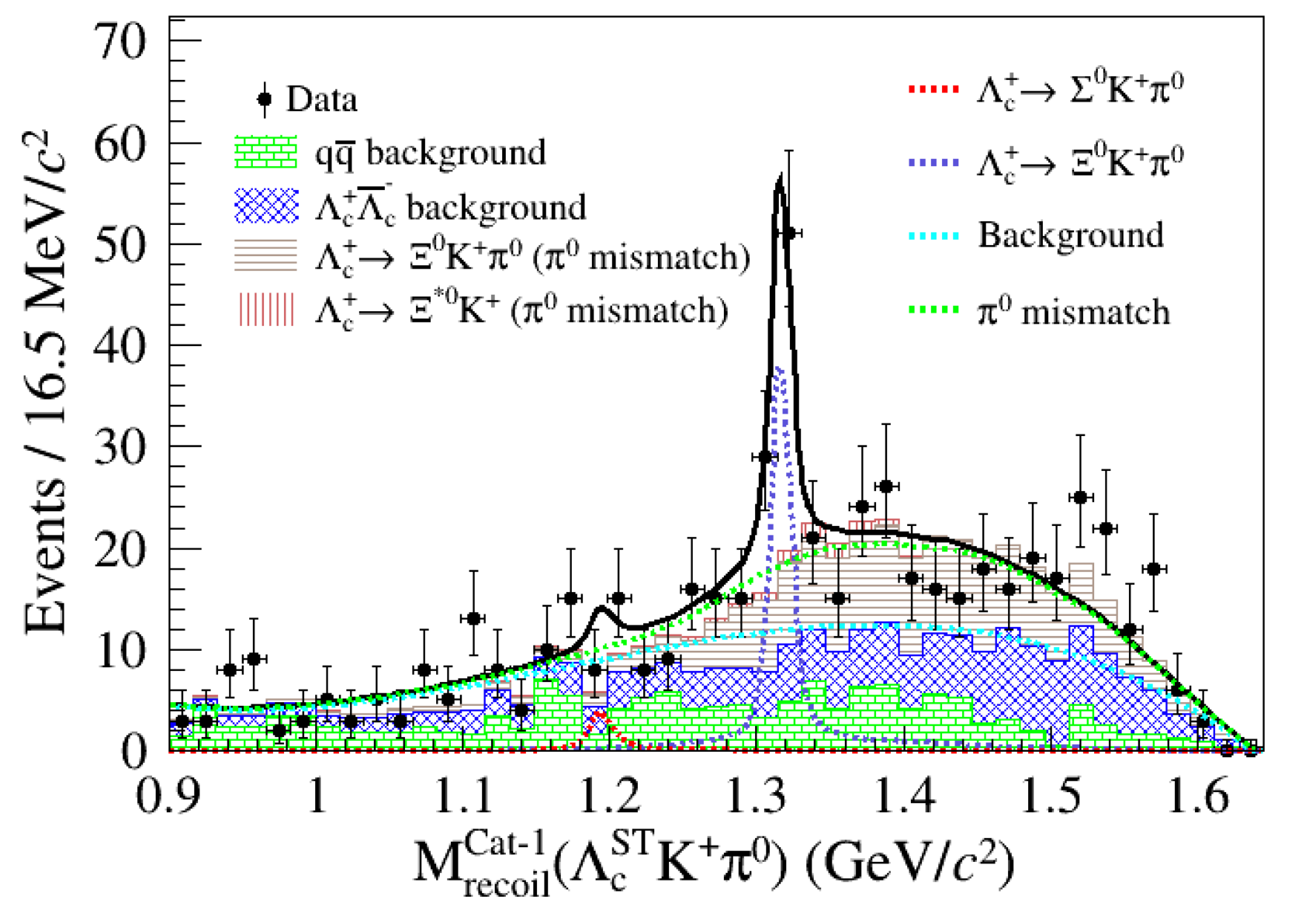}\label{fig:sigma0kpi0_cat1}}
  \subfigure[]
   { \includegraphics[width=0.319\textwidth,height=0.226\textwidth]{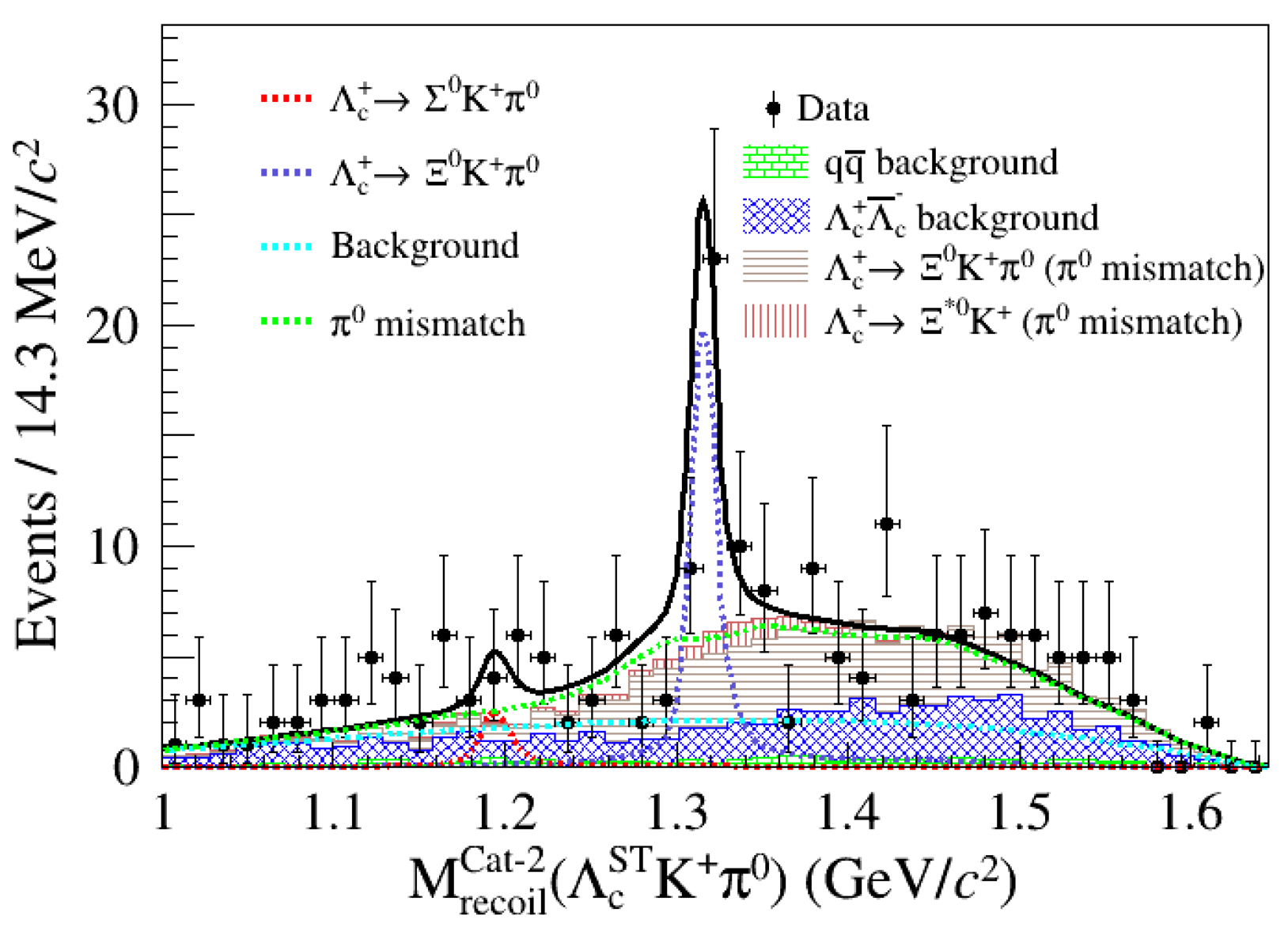}\label{fig:sigma0kpi0_cat2}}      
    \subfigure[]      
  {  \includegraphics[width=0.319\textwidth,height=0.226\textwidth]{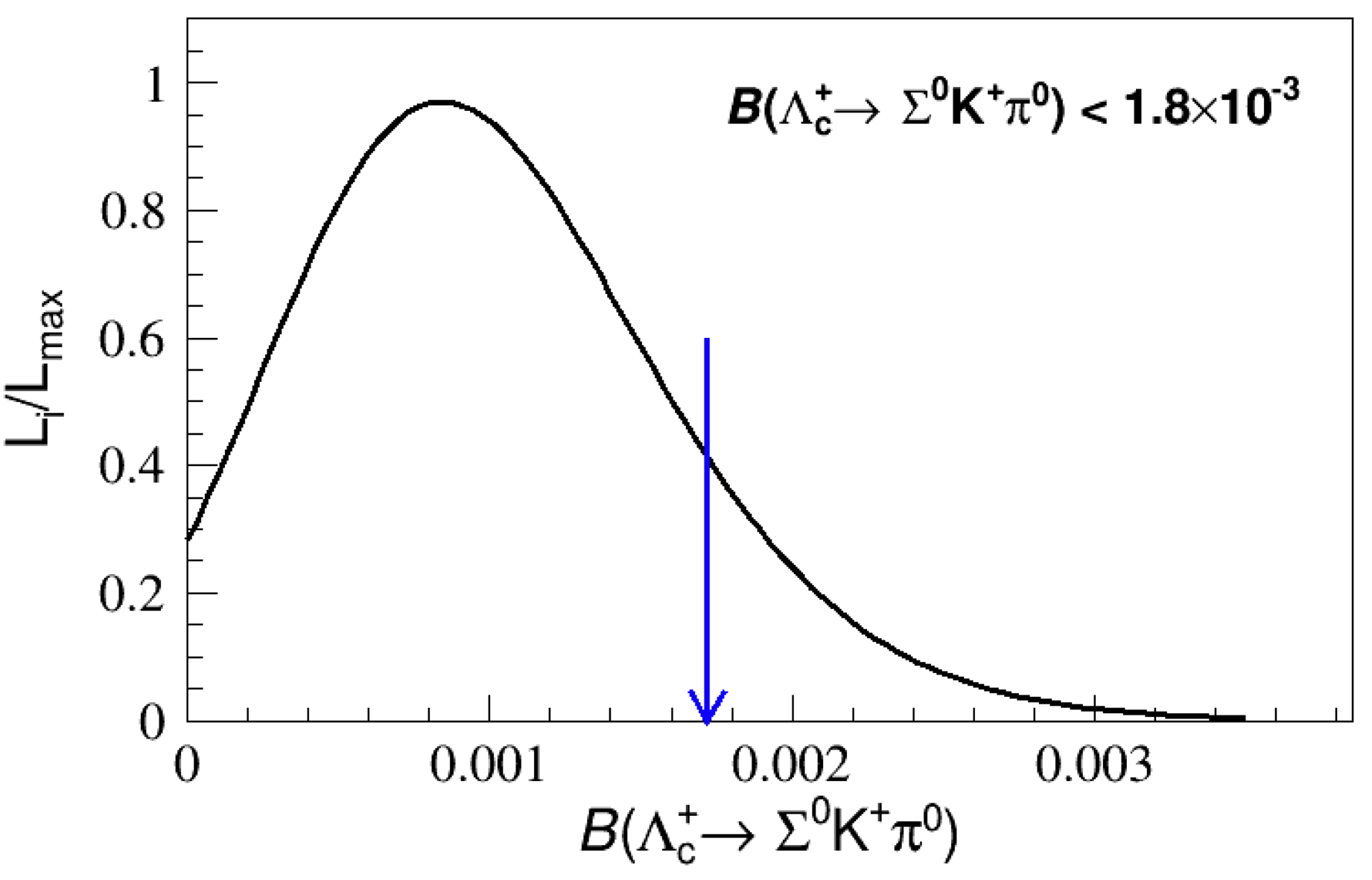}\label{fig:noComfit_uplimit_1} }             
    \caption{Fit projections on (a) $\MrecPKPI$, (b) $\MrecPKPIB$
    distributions and (c) Likelihood distributions over the branching
    fractions of {\LamCSigKPi}.  The black curve denotes the fit
    result with systematic uncertainty. The blue arrows show the
    results corresponding to 90\% confidence
    level.}\label{fig:sigma0kpi0} \end{figure*}

 \begin{figure*}[!htp]
 \centering
 \subfigure[]{\includegraphics[width=0.319\textwidth,height=0.226\textwidth]{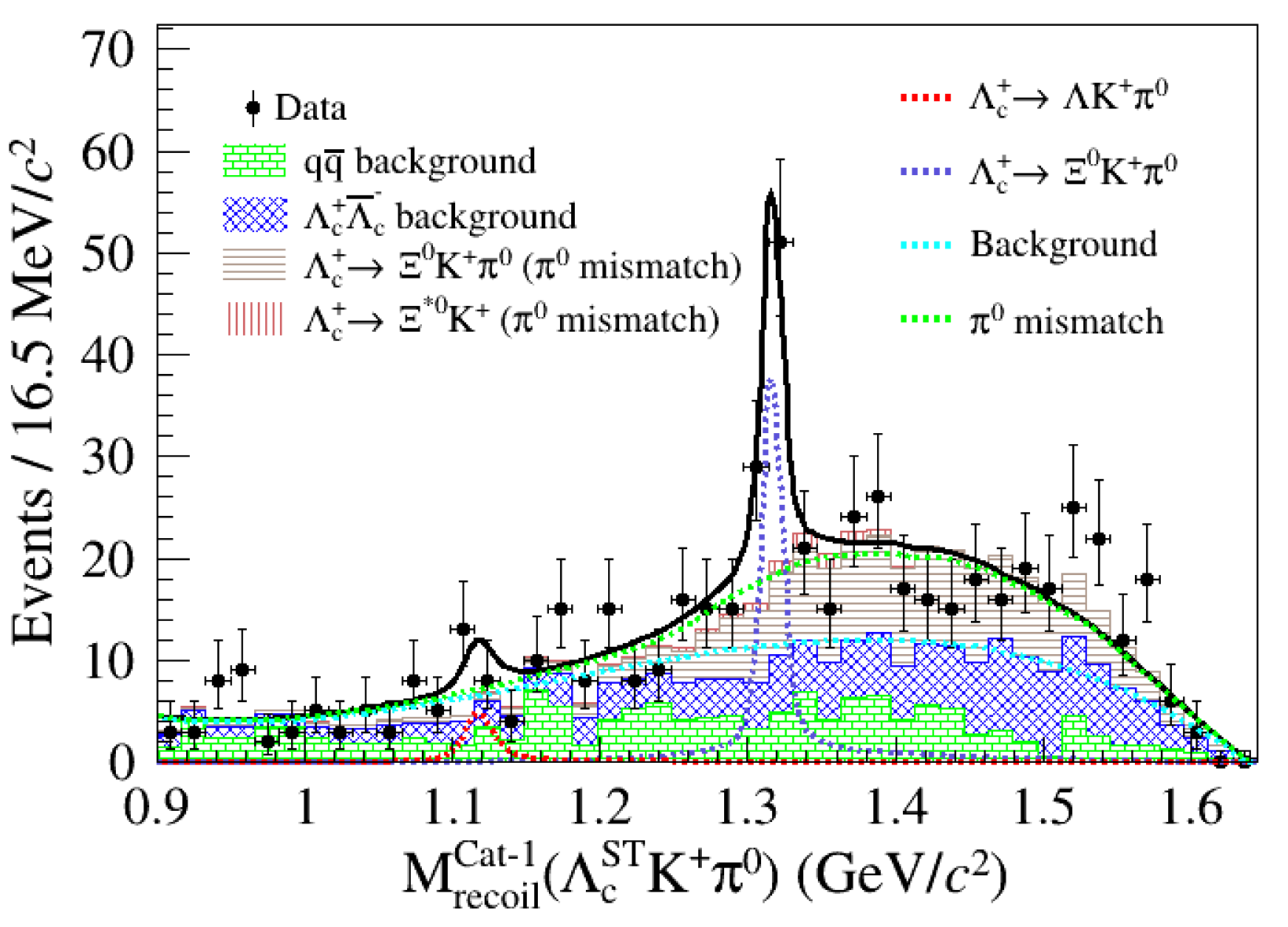}\label{fig:lambdakpi0_cat1}}
  \subfigure[]
   { \includegraphics[width=0.319\textwidth,height=0.226\textwidth]{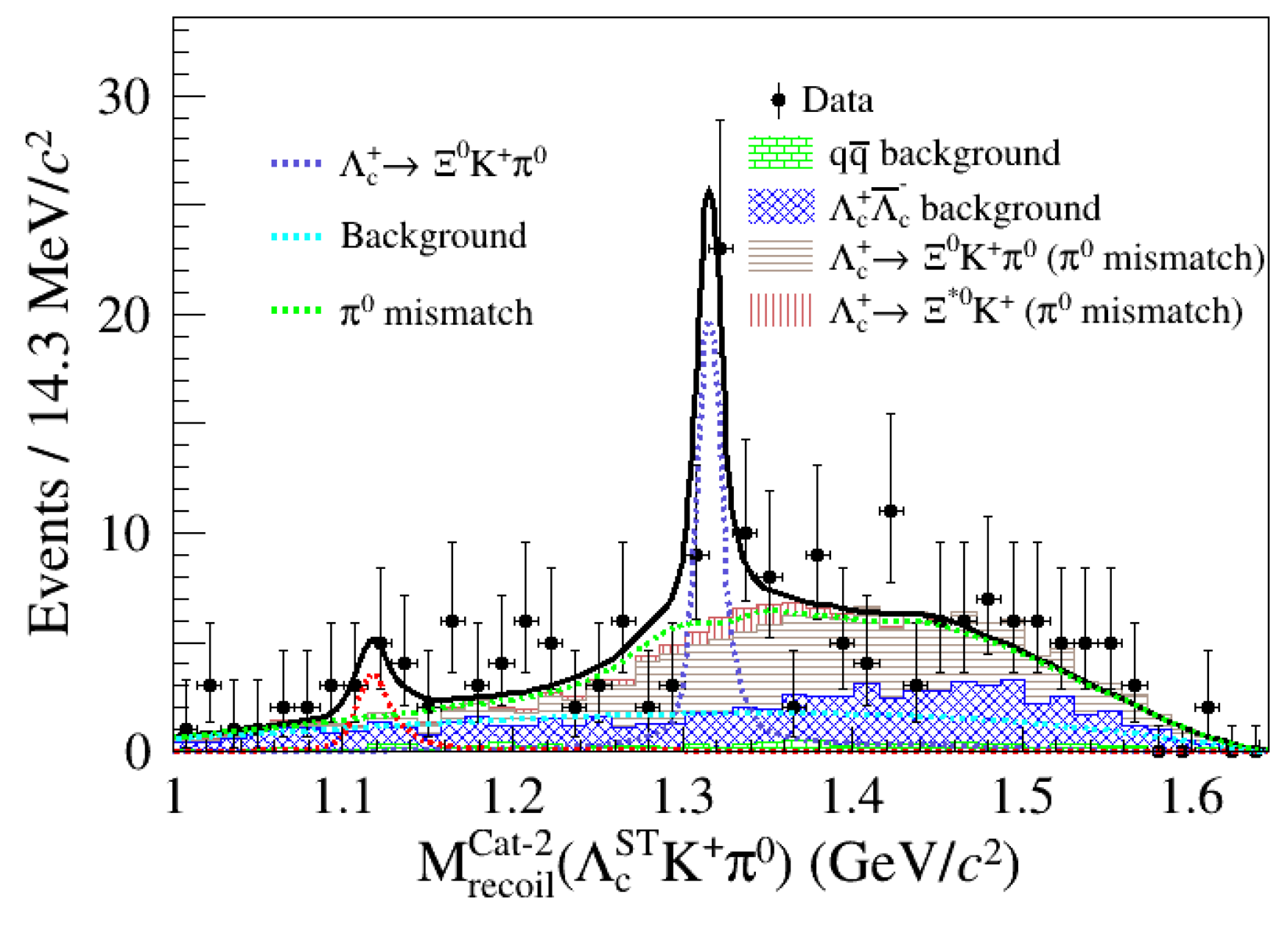}\label{fig:lambdakpi0_cat2}}      
    \subfigure[]      
  {  \includegraphics[width=0.319\textwidth,height=0.226\textwidth]{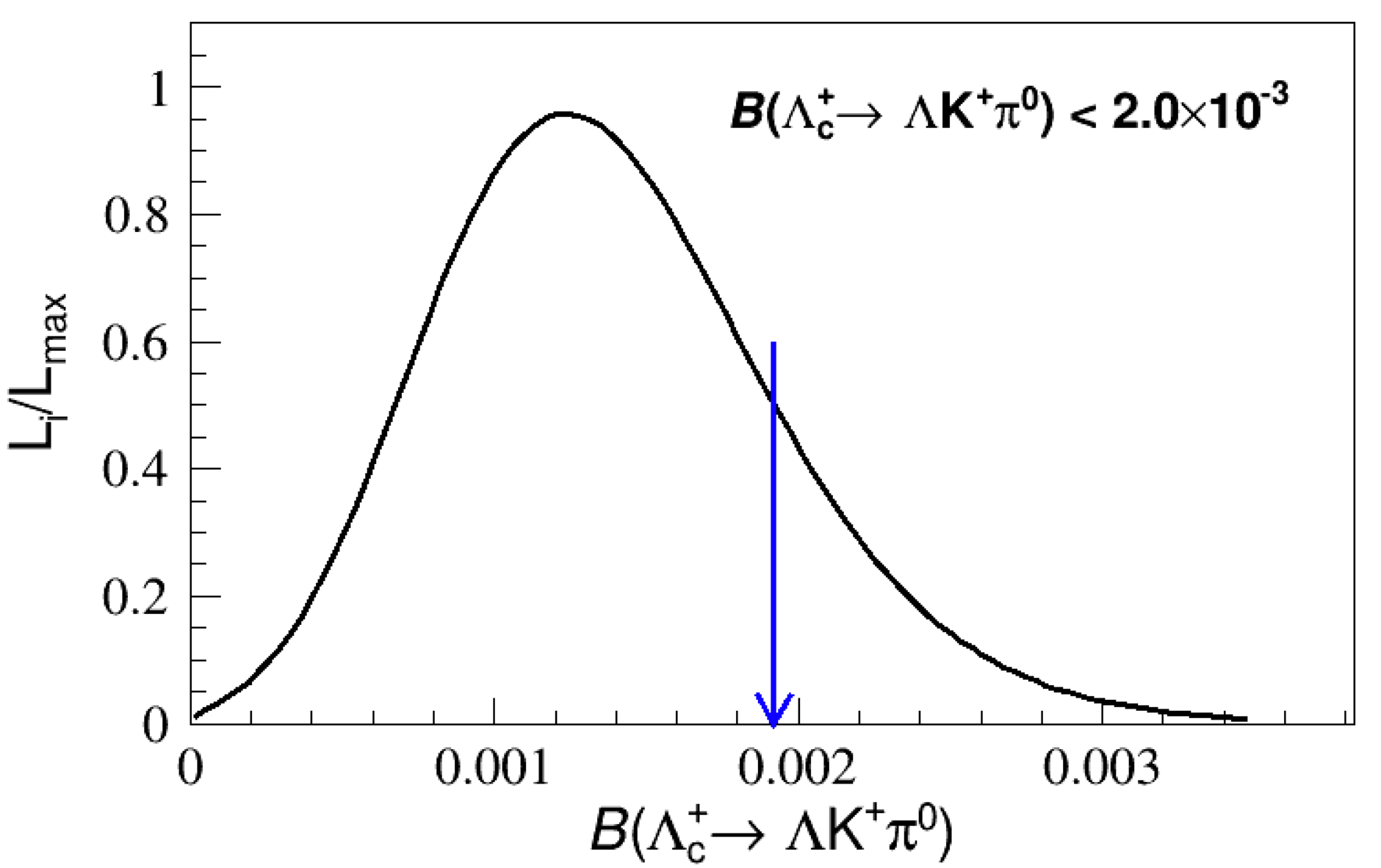}\label{fig:noComfit_uplimit_2} }             
    \caption{Fit projections on (a) $\MrecPKPI$, (b) $\MrecPKPIB$
    distributions and (c) Likelihood distributions over the branching
    fractions of $\LamC \to \Lambda K^{+}\pi^{0}$.
    }\label{fig:lambdakpi0}
\end{figure*}

For the three-body decays $\LamCNKPi$, $\LamCSigKPi$ and $\Lambda
K^{+}\pi^{0}$, simultaneous fits are made to the $\MrecPKPI$ and
$\MrecPKPIB$ distributions as shown in Figs.~\ref{fig:nkpi0},
\ref{fig:sigma0kpi0} and \ref{fig:lambdakpi0}.  
Since no significant signals are observed, the frequentist method is used to determine the upper limits on the branching fractions of
these decays~\cite{method}. 
The upper limits, at a 90\% confidence level, are determined by 
integrating the likelihood curves, which are obtained by scanning over 
the branching fraction.

For the three-body
decay $\LamCNKPi$, the signal yield is $10 \pm 6$ events and DT efficiencies in Cat-1 are summarized in Table~\ref{case1-nkpi0}. 
The upper limit on the branching fraction of this decay at the 90$\%$ confidence level is ${\mathcal B}(\LamCNKPi)
< 7.1\times10^{-4}$.
For the three-body decays $\LamCSigKPi$ and
$\Lambda K^{+}\pi^{0}$, the signal yields of $\LamCSigKPi$ are $7 \pm 5$ and $6 \pm 4$ events for Cat-1 and Cat-2, respectively. The signal yields of $\LamCLamKPi$ are $10 \pm 4$ and $10 \pm 4$ events for Cat-1 and Cat-2, respectively. 
The DT efficiencies in are summarized in
Tables~\ref{case1-sigma0kpi0} and \ref{case1-lambdakpi0}, and the
upper limits are ${\mathcal B}(\LamCSigKPi) < 1.8\times10^{-3}$ and
${\mathcal B}(\LamCLamKPi) < 2.0\times10^{-3}$, respectively. The
black solid curves in Figs.~\ref{fig:noComfit_uplimit_3}, \ref{fig:noComfit_uplimit_1} and
\ref{fig:noComfit_uplimit_2} show the
resulting likelihood distributions for these three decays.

\section{Systematic uncertainties}

  \begin{table*}[!htbp]
    \begin{center}
    \caption{Systematic uncertainties (\%). }
    \label{tab:sys-sum}
    \begin{tabular}{ l |cc |cc |cc |cc |c}
      \hline
      \hline
	  \multirow{2}*{Sources} & \multicolumn{2}{c|}{\LamCXistarK} &  \multicolumn{2}{c|}{\LamCXiKPi}  & \multicolumn{2}{c|}{\LamCSigKPi} &\multicolumn{2}{c|}{\LamCLamKPi} & \multicolumn{1}{c}{\LamCNKPi} \\ \cline{2-10} 
	            &  Cat-1 & Cat-2 &  Cat-1 & Cat-2 &  Cat-1 & Cat-2 &  Cat-1 & Cat-2 &  Cat-1  \\
      \hline
	    No extra charged track & \multicolumn{9}{c}{2.20} \\ 
	    $K^{+}$ tracking  & \multicolumn{9}{c}{1.00} \\   
	    $K^{+}$ PID & \multicolumn{9}{c}{1.00} \\ 
	    $\pi^{0}$ reconstruction & \multicolumn{9}{c}{1.00} \\ 
	    ST yield &  \multicolumn{9}{c}{0.50} \\
	    Background PDF & 1.17 & 1.17 & 3.70 & 3.70  & 4.18 & 4.18 & 3.68 & 3.68 & 3.99 \\
	    $\Lambda$ reconstruction & ---  & 0.60 & --- & 0.60 & --- & 0.60 & ---  & 0.60 & --- \\
	    $\pi^0$ selection & 2.23 & 2.23 & 5.23  & 5.23  &2.86  & 2.68 & 3.56 & 3.42 & 4.29 \\
	    Reweighting  & --- & --- & 4.95 & 2.43 & ---  & --- & ---  & --- & --- \\
     \hline
            Sum  & 3.80 & 3.85 & 8.58 &8.60 & 5.81 & 5.76 & 5.86&5.81&6.51\\
      \hline\hline
\end{tabular}
  \end{center}
\end{table*}

The systematic uncertainties for the branching fraction measurements
include those associated with the ST yields ($N^{\rm ST}_{i}$),
detection efficiencies of the ST \LamCBar{} ($\epsilon^{\rm ST}_{i}$),
detection efficiencies of the DT events ($\epsilon^{\rm DT}_{i}$) and
signal yield ($N_{\mathrm{sig}}$).  Since Eq.~\ref{eq:br1} contains a
ratio of ST and DT efficiencies, any systematic uncertainty on the tag
side is canceled to first order. Each of them is evaluated relative to
the measured branching fraction.

The numbers of charged tracks are required to be only one (Cat-1) or
three (Cat-2) without any extra charged tracks. This difference
between data and MC simulation for this selection is studied with
control sample of $\LamC\to pK^-\pi^+$ (the other $\LamCB$ goes to the
10
tag modes). The uncertainty on the requirement of the number of
charged tracks is assigned to be 2.2\%, which is denoted as ``No extra
charged track'' in Table~\ref{tab:sys-sum}.

The tracking efficiencies for $K^{+}$ as a function of transverse momentum
has been studied with the process $J/\psi\to K^{0}_{S}K^{+}\pi^{-} \to
K^{+}\pi^{+}\pi^{-}\pi^{-}$. The efficiency differences between data and
MC simulation are both 1\% for $K^+$ tracking and PID
efficiencies, which are taken as the systematic
uncertainties~\cite{kaon_un}. The uncertainty associated with the $\pi^{0}$ reconstruction is assigned to be 1.0\%, studied with control sample of $J/\psi\to \pi^{+}\pi^{-}\pi^{0}$~\cite{pi0_un}.

The uncertainties in the total ST yields are 0.5\%~\cite{STyield_un}, which arise from the statistical uncertainty and fitting
strategy of extracting ST yields. 

The uncertainties of the background and fits of the distributions of
$\MrecPKPI$ and $\MrecPKPIB$ arise from the sideband
range and background estimation. First, the uncertainty for the data
sideband and $q\bar{q}$ background estimation is derived from the
statistical uncertainty of the background estimate in the $M_{\rm BC}$
sideband region and the variation of the sideband range. Second, the
background shape is changed from a polynomial function to an ARGUS
function. The differences between the new and nominal results are taken
as the systematic uncertainties.

The uncertainty due to the $\Lambda$ reconstruction efficiency, which
is estimated by a weighted root-mean-square of the statistical
uncertainties for different ($p$, $\cos\theta$) intervals, is assigned
to be 0.6\% using a sample of $J/\psi\to \Bar p K^{+}\Lambda$
decays~\cite{z24}.

The uncertainty due to the $\pi^0$ selection combines two sources.  The first source is originated from $\pi^0$ selection with the minimum
kinematic-fit $\chi^2$, which is studied with control sample of $\LamC \to
\Sigma^{+}\omega$ and $\LamC \to \Sigma^{+}\pi^{0}$. The difference
between data and MC simulation is 1\%.
The second one is due to the $\pi^0$ mismatch component, which is
evaluated using simulation samples with an alternative $\pi^0$
matching approach, where the $\pi^0$ candidate closest to the true
$\pi^0$ is chosen to be the reconstructed $\pi^0$ candidate. The fit
is performed on the simulation samples, using both the nominal and
alternative $\pi^0$ matching, and the difference in the fitted
branching fraction between them is taken as the systematic
uncertainty.
 
The uncertainty in reweighting the $\LamCXiKPi$ simulation samples is
derived using different 2D variables and different choices of 2D
binning, for instance 6 bins $\times$ 6 bins to 7 bins $\times$ 7
bins. The uncertainty is determined in a similar manner as the
uncertainty from the background shape.

The total systematic uncertainty for the two strategies is obtained by
taking the quadratic sum of the individual values.  Some systematic
uncertainties are the same between two strategies, e.g. tracking and
PID of the charged kaon, so they are assumed to be fully correlated in
the combination. Other systematic uncertainties are independent
between two strategies, e.g. $\Lambda$ reconstruction, background
estimation in the fits on the distributions of $\MrecPKPI$ and
$\MrecPKPIB$, and they are treated without any correlation in the
combination. The different systematic sources are listed in
Table~\ref{tab:sys-sum}, where the two categories have been separated
explicitly.

	\begin{table*}[!htbp]
  \begin{center}
	  \caption{The comparison between the measurement and theoretical predictions ($\times {10^{ - 3}}$). The first and the second uncertainties are statistical and systematic, respectively. }\label{tab:theory_prediction}
  \renewcommand\arraystretch{2.0}
  \setlength{\tabcolsep}{1.0 pt}
    \scalebox{0.9}{
    \begin{tabular}{ l | c | c | c | c | c}
      \hline
      \hline
           &  \LamCXistarK   &  $\LamCXiKPi$ &  \LamCSigKPi  &  \LamCLamKPi &  \LamCNKPi  \\
      \hline
      This measurement & $ 5.99\pm1.04\pm 0.32$ &  $7.79\pm1.46 \pm0.95$ & $ < {1.8}$ & $ < {2.0}$ & $ < {0.71}$  \\
      \hline
       K. K. Sharma {\it et al .}~\cite{z16} &$-$ & $45\pm8$ & $ {1.2\pm0.3}$ & ${4.5\pm 0.8}$ &  $0.05\pm 0.005$\ \\
      \hline
       Jian-Yong Cen {\it et al .}~\cite{z30}  &$-$ & $32\pm6$  & $ {0.7\pm0.2}$ & ${3.5\pm 0.6}$ &  $0.05\pm 0.006$\ \\
      \hline
       \br(previous results)~\cite{tobepu} &${5.02\pm 0.99\pm0.31}$ &$-$  & $-$ & $-$ &  $-$\\
       \hline\hline
    \end{tabular}
    	     }

  \end{center}
\end{table*}

\section{Summary} With ${6.1}~\ifb{}$ of $\ee{}$ collision data
collected at eleven CM energy points between 4.600 and 4.840~GeV with
the BESIII detector at BEPCII, the CF decays $\LamCXiKPi{}$ and
$\LamCXistarK{}$ are observed with significances of 8.6$\sigma$ and
6.9$\sigma$, respectively.  The branching fraction of $\LamCXiKPi{}$
is measured to be $(7.79\pm1.46 \pm0.95)\times 10^{-3}$, where the
first and second uncertainties are statistical and systematic,
respectively. It is smaller than the theoretical predictions
$(4.5\pm0.8) \times {10^{ - 2}}$~\cite{z16} and $(3.2\pm0.6) \times
{10^{-2}}$~\cite{z30}. Comparisons of theory and experiment are shown
in Table~\ref{tab:theory_prediction}. The branching fraction of
$\LamCXistarK{}$ is $(5.99\pm1.04\pm0.32)\times 10^{-3}$, which is
consistent with the previous result $(5.02\pm0.99\pm0.31)\times
10^{-3}$~\cite{tobepu}.

The upper limits on the branching fractions at the 90$\%$ confidence
level of the SCS or DCS decays \LamCNKPi{}, $\Sigma^{0}K^{+}\pi^{0}$
and $\Lambda K^{+}\pi^{0}$ are $7.1 \times 10^{-4}$, $1.8\times
10^{-3}$ and $2.0 \times 10^{-3}$, respectively. The upper limit of
the branching fraction of $\LamCLamKPi{}$ is incompatible with the
theory predictions~\cite{z16,z30}. The upper limits of the branching
fractions of $\LamCSigKPi{}$ and \LamCNKPi are consistent with
the theoretical predictions~\cite{z16,z30}. These results are
essential for the understanding of the dynamics in the charmed baryon
decays.

\section{Acknowledgement}

The BESIII collaboration thanks the staff of BEPCII$~$ and the IHEP computing center for their strong support. This work is supported in part by National Key R\&D Program of China under Contracts Nos. 2020YFA0406400, 2020YFA0406300; National Natural Science Foundation of China (NSFC) under Contracts Nos. 11635010, 11735014, 11835012, 11935015, 11935016, 11935018, 11961141012, 12022510, 12025502, 12035009, 12035013, 12192260, 12192261, 12192262, 12192263, 12192264, 12192265, 12221005, 12225509, 12235017, 12005311; the Fundamental Research Funds for the Central Universities, Sun Yat-sen University, University of Science and Technology of China; 100 Talents Program of Sun Yat-sen University; the Chinese Academy of Sciences (CAS) Large-Scale Scientific Facility Program; Joint Large-Scale Scientific Facility Funds of the NSFC and CAS under Contract No. U1832207; the CAS Center for Excellence in Particle Physics (CCEPP); 100 Talents Program of CAS; The Institute of Nuclear and Particle Physics (INPAC) and Shanghai Key Laboratory for Particle Physics and Cosmology; ERC under Contract No. 758462; European Union's Horizon 2020 research and innovation programme under Marie Sklodowska-Curie grant agreement under Contract No. 894790; German Research Foundation DFG under Contracts Nos. 443159800, Collaborative Research Center CRC 1044, GRK 2149; Istituto Nazionale diFisica Nucleare, Italy; Ministry of Development of Turkey under Contract No. DPT2006K-120470; National Science and Technology fund; National Science Research and Innovation Fund (NSRF) via the Program Management Unit for Human Resources \& Institutional Development, Research and Innovation under Contract No. B16F640076; Olle Engkvist Foundation under Contract No. 200-0605; STFC (United Kingdom); Suranaree University of Technology (SUT), Thailand Science Research and Innovation (TSRI), and National Science Research and Innovation Fund (NSRF) under Contract No. 160355; The Royal Society, UK under Contracts Nos. DH140054, DH160214; The Swedish Research Council; U. S. Department of Energy under Contract No. DE-FG02-05ER41374.

\end{document}

%% file: authorlist_2023-3-6.tex
\author{
M.~Ablikim$^{1}$, M.~N.~Achasov$^{5,c}$, P.~Adlarson$^{75}$, O.~Afedulidis$^{4}$, X.~C.~Ai$^{80}$, R.~Aliberti$^{36}$, A.~Amoroso$^{74A,74C}$, M.~R.~An$^{40}$, Q.~An$^{71,58,a}$, Y.~Bai$^{57}$
% ...
\\
\vspace{0.2cm}
(BESIII Collaboration)\\
\vspace{0.2cm} {\it
$^{1}$ Institute of High Energy Physics, Beijing 100049, People's Republic of China\\
% ...
$^{4}$ Bochum Ruhr-University, D-44780 Bochum, Germany\\
$^{5}$ Budker Institute of Nuclear Physics SB RAS (BINP), Novosibirsk 630090, Russia\\
% ...
$^{36}$ Johannes Gutenberg University of Mainz, Johann-Joachim-Becher-Weg 45, D-55099 Mainz, Germany\\
% ...
$^{40}$ Liaoning Normal University, Dalian 116029, People's Republic of China\\
% ...
$^{57}$ Southeast University, Nanjing 211100, People's Republic of China\\
$^{58}$ State Key Laboratory of Particle Detection and Electronics, Beijing 100049, Hefei 230026, People's Republic of China\\
% ...
$^{71}$ University of Science and Technology of China, Hefei 230026, People's Republic of China\\
% ...
$^{74}$ University of Turin and INFN, (A)University of Turin, I-10125, Turin, Italy; (B)University of Eastern Piedmont, I-15121, Alessandria, Italy; (C)INFN, I-10125, Turin, Italy\\
$^{75}$ Uppsala University, Box 516, SE-75120 Uppsala, Sweden\\
% ...
$^{80}$ Zhengzhou University, Zhengzhou 450001, People's Republic of China\\
\vspace{0.2cm}
$^{a}$ Also at the Moscow Institute of Physics and Technology, Moscow 141700, Russia\\
% ...
$^{c}$ Also at the NRC "Kurchatov Institute", PNPI, 188300, Gatchina, Russia\\
% ...
}
}